\documentclass[journal]{IEEEtran}

\usepackage{amsmath,bm}
\usepackage{array}
\usepackage{xcolor}
\usepackage{amssymb,amsfonts}
\usepackage{booktabs}
\usepackage{scalerel}
\usepackage[all,arc]{xy}
\usepackage{enumerate}
\usepackage{mathrsfs}
\usepackage[pass]{geometry}
\usepackage{titletoc}
\usepackage[noend]{algpseudocode}
\usepackage{graphicx}
\usepackage[caption=false,font=footnotesize]{subfig}
\usepackage{balance}
\usepackage{algorithm}
\usepackage{lipsum}
\usepackage{dblfloatfix} 
\usepackage{cite}
\usepackage{enumitem}

\usepackage[version=4]{mhchem}
\usepackage{chemfig}
\usepackage[strings]{underscore}
\usepackage[caption=false]{subfig}
\usepackage{siunitx}
\usepackage{scalerel}
\usepackage{multirow}
\usepackage{colortbl}
\usepackage{upgreek}
\usepackage[inkscapearea=page]{svg}
\usepackage{fixltx2e,amsmath}
\usepackage{nicematrix}
\MakeRobust{\eqref}
\graphicspath{{figure/}}

\newcommand{\thickbar}[1]{\mathbf{\bar{\text{$#1$}}}}

\makeatletter
\newcommand*\bigcdot{\mathpalette\bigcdot@{.5}}
\newcommand*\bigcdot@[2]{\mathbin{\vcenter{\hbox{\scalebox{#2}{$\m@th#1\bullet$}}}}}
\makeatother

\makeatletter
\renewcommand\footnoterule{
  \ifnum\c@page=2 
    \kern-3\p@
    \hrule \@width .4\columnwidth
    \kern 2.6\p@ 
  \fi
}
\makeatother

\DeclareSIUnit\rpm{rpm}

\begin{document}

\title{Machine Learning for the Control and Monitoring of Electric Machine Drives: Advances and Trends}

%
\author{Shen Zhang, Oliver Wallscheid, and Mario Porrmann
\thanks{Shen Zhang was with the School of Electrical and Computer Engineering, Georgia Institute of Technology, Atlanta, GA 30332, USA. He is now with Joby Aviation, San Carlos, CA 94070, USA (e-mail: shenzhang@gatech.edu).}
\thanks{Oliver Wallscheid is with the Department of Automatic Control, Paderborn University, 33098 Paderborn, Germany (e-mail: oliver.wallscheid@uni-paderborn.de).}

\thanks{Mario Porrmann is with the Institute of Computer Science, Osnabr{\"u}eck University, Wachsbleiche 27, 49090 Osnabr{\"u}ck, Germany (email: mario.porrmann@uni-osnabrueck.de).}
}
%

%


\maketitle
\thispagestyle{plain}
\pagestyle{plain}
\begin{abstract}
This review paper systematically summarizes the existing literature on utilizing machine learning (ML) techniques for the control and monitoring of electric machine drives. It is anticipated that with the rapid progress in learning algorithms and specialized embedded hardware platforms, machine learning-based data-driven approaches will become standard tools for the automated high-performance control and monitoring of electric drives. Additionally, this paper also provides some outlook toward promoting its widespread application in the industry with a focus on deploying ML algorithms onto embedded system-on-chip (SoC) field-programmable gate array (FPGA) devices. 
\end{abstract}

\section{Introduction} \label{sec:Introduction}
The motor control community is well-informed on the boom of machine learning (ML) after the modern back-propagation paper was first published in 1986 \cite{rumelhart1986learning}, which is evident by the work that appeared three years later on training a neural network offline to mimic the behavior of hysteresis current controllers in a three-phase PWM inverter \cite{harashima1989application}. This work is later followed by a series of pioneering efforts in the early 1990s on general voltage-fed AC machines \cite{buhl1991design, lin1993power}, induction machines \cite{ben1993identification, mir1994fuzzy, wishart1995identification, kung1995adaptive, toh1994flux, theocharis1994neural, simoes1995neural, peng1994robust, ben1995motor, sousa1995fuzzy, mehrotra1996development}, DC machines \cite{weerasooriya1991identification, rahman1997online}, synchronous machines \cite{tsai1995development}, and switched reluctance machines \cite{reay1995switched}. In addition to the broad interest in applying ML to motor drive control, such technologies, especially concerning classification or regression techniques, have also found their presence in the condition monitoring and fault diagnosis on various types of electric machines \cite{goode1995using1, filippetti1998ai, filippetti2000recent, tallam2002self, awadallah2003application, huang2007detection, bouzid2008effective, mohagheghi2009condition}. 

Around that time, the frontier of power electronics gradually advanced with the advent of ML models such as neural networks, which have emerged as the most important area for complex system identification, control, and estimation in power electronics and motor drives \cite{cirrincione2017book}. However, it was also concluded  that ``in spite of the technology advancement, currently, industrial applications of neural networks in power electronics appear to be very few'' \cite{bose2007neural}. 

While ML applications always targeted the fastest available hardware platforms, especially focusing on (massively) parallel architectures, many existing ML implementations in electric machine drives were based on slow and sequentially executed digital signal processors (DSP) prior to the deep learning era, although in some cases multiple DSPs were also used to increase the execution speed. Embedded platforms such as field-programmable gate arrays (FPGA) that excel at parallel processing, were not matured technologies at the time and had limited use. 

It is worth noting that hardware limitations are still the main bottleneck for ML applications in electric machine drives even today. This remains a major problem particularly in the industrial world due to the high-frequency update rates\footnote{While ML algorithms for real-time video processing only require to run at a couple of ten hertz, typical motor control update frequencies are in the range of ten kilohertz, so multiple orders of magnitude higher.} of ML algorithms required for motor drive online applications in combination with cost-oriented, computationally-constrained embedded hardware. This hardware constraint has further impeded the deployment of ML algorithms in machine drives and resulted in insufficient performance in their identification and control. It is envisioned in \cite{bose2012global} that with the present trend of FPGA development and as the ML technology matures, ``intelligent control and estimation (particularly based on neural networks) will find increasing acceptance in power electronics, particularly in the robust control of drives'' \cite{bose2020power}, and they are expected to have widespread applications in the industry \cite{bose2017artificial, bose2020artificial, zhao2020overview}. 
\begin{figure}[!t]
\centering
\includegraphics[width=3.4in]{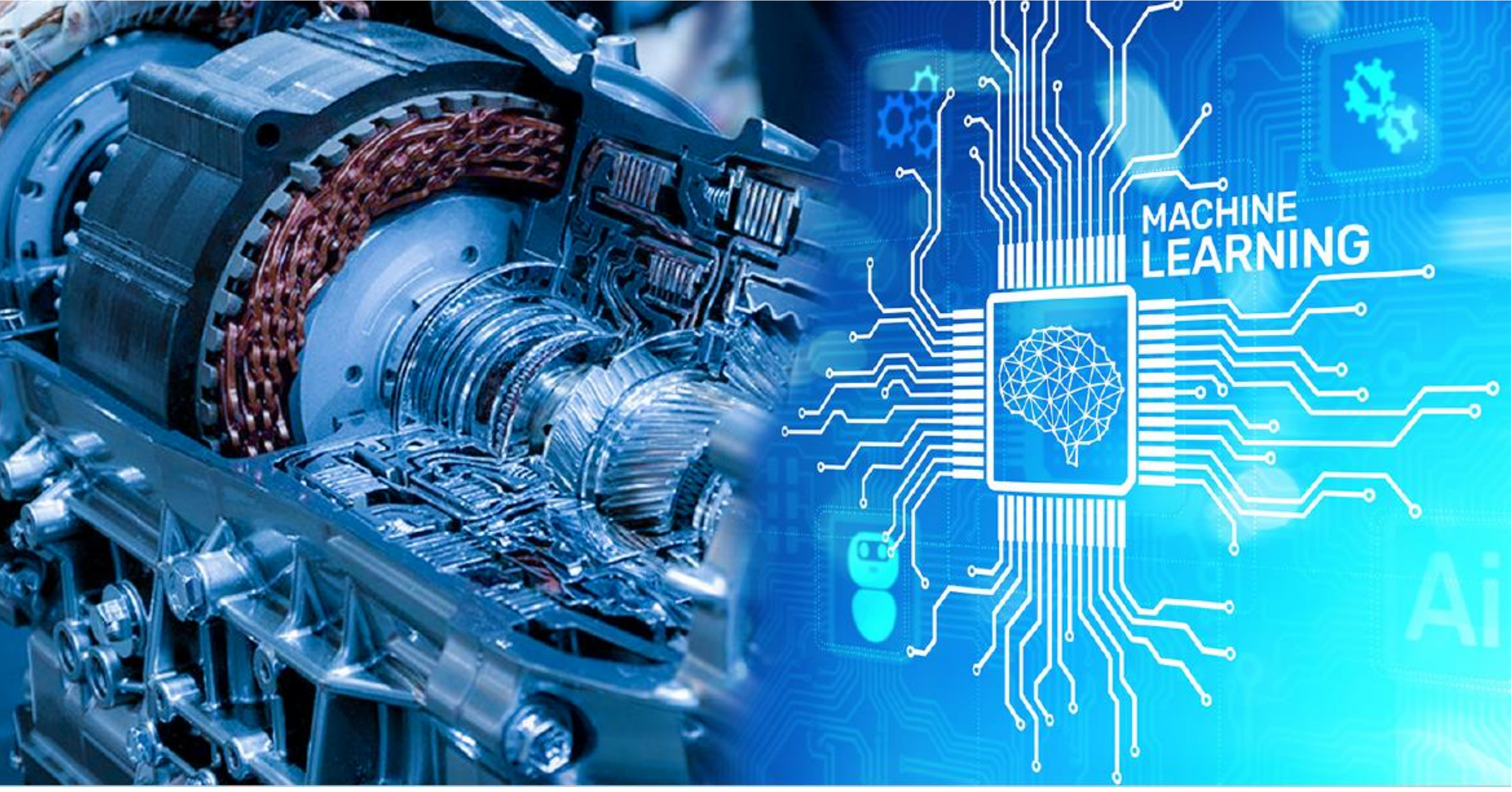}
\caption{A vision of future electric machine drives powered by machine learning.}
\label{fig:motor_ml}
\end{figure}

The past decade has marked an incredibly fast-paced and innovative period in the history of ML \cite{deep_learning_decade}. Spurred by the development of ever-more-powerful computing platforms and the increased availability of big data, ML has successfully tackled many previously intractable problems, especially in computer vision with the convolutional neural network (CNN) \cite{krizhevsky2012imagenet} and in natural language processing with the Transformer architecture \cite{vaswani2017attention}. ML has also been applied to and is in the process of transforming many real-world applications, including entertainment, healthcare, fraud detection, virtual assistants, and autonomous vehicles. Hardware platforms including GPUs and FPGA fabric can also achieve very good parallel computing performance with architecture customization \cite{monmasson2021system}, which is intrinsically well-suited for the parallel characteristics inherent in neural networks and hence their widespread applications in power electronics and motor drives.

However, the entire field of electric machine drives has remained pretty much silent on the resurgence of ML over the past decade, when compared with its continued success and widespread application in condition monitoring \cite{liu2014position, kumar2020topological, bengherbia2020fpga, zhang2020deep, nath2020role, lee2020temperature, cai2021temperature}, design optimization \cite{zhang2018efficient, zhang2018visualization, khan2019deep, doi2019multi, sasaki2019topology, zhang2019visualization, guillod2020artificial, barmada2020deep, khan2020efficiency, guillod2020brute, li2020magnet, hao2020optimization, parekh2021deep, sato2021data, barmada2021deep, sasaki2021explainable, li2021machine, saha2021machine, you2020multi, gabdullin2021towards}, and manufacturing \cite{mayr2018application, mayr2021towards} of various types of electric machines. It was not until in the last few years that research efforts have begun to gradually catch up with the trend \cite{schenke2019controller, traue2020toward, balakrishna2021gym, hanke2020data1, hanke2020data2, schenke2021deep, book2021transferring, schindler2019comparison, bhattacharjee2020advanced, el2020adaptive, alharkan2021optimal}. It is anticipated that with the rapid progress in ML models and embedded systems, the data-driven approach will become increasingly popular for the high-performance control of electric machine drives, as envisioned in Fig. \ref{fig:motor_ml}. While most ML models require some complex offline training processes, the online inference process can be made relatively simple through various pruning and quantization methods \cite{monmasson2021system}, such that the groups of parameters with insignificant impact for the artificial neural network's (ANN) input-output characteristics are removed and the numeric precision of the weights is reduced, resulting in reduced model size and faster computation at the cost of minimal reductions in predictive accuracy \cite{seng2021embedded}.

\subsection{Machine Learning Algorithms}
\begin{figure}[!t]
\centering
\includegraphics[width=3.4in]{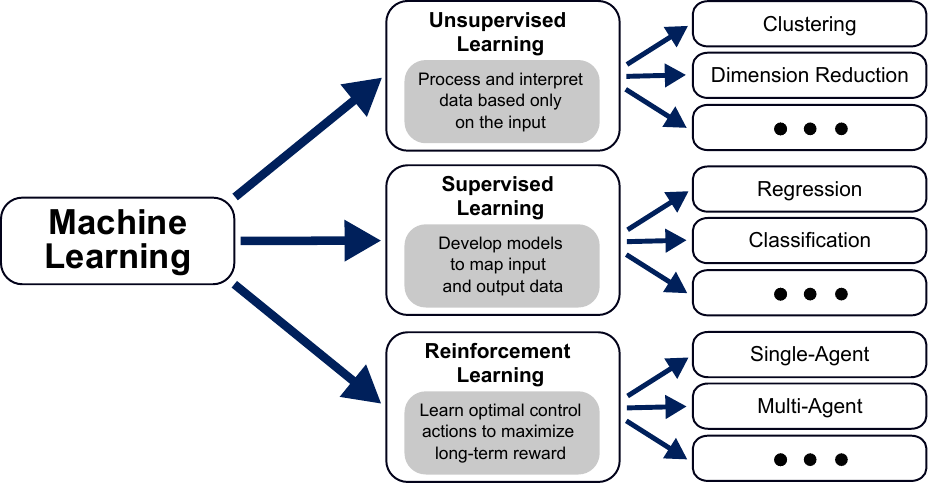}
\caption{Disciplines of machine learning \cite{KSWW2020}.}
\label{fig:ml_disciplines}
\end{figure}

There are three main classes of ML: supervised learning, unsupervised learning, and reinforcement learning. A brief description of each class and its classic applications are illustrated in Fig. \ref{fig:ml_disciplines} and in the paragraphs below.
\subsubsection{Supervised Learning}
Supervised learning is the ML task of developing models to map input and output data. The term ``supervised'' refers to the use of labeled data to train models for classification or regression problems. Some of the most widely used supervised learning algorithms include ordinary least squares, ANN, support vector machines, etc. As will be discussed in later sections, these algorithms have been extensively applied to the control and monitoring of electric machine drives, as well as the estimation of models or model parameters associated with the drives.
\subsubsection{Unsupervised Learning}
Unsupervised learning is a type of algorithm that processes and interprets data based solely on input. Classical tasks in unsupervised learning include clustering, dimensionality reduction, and anomaly detection. In contrast to supervised learning, unsupervised learning models work on their own to discover the inherent structure of unlabeled data. Unsupervised learning can also be used as an auxiliary pre-processing step in order to apply feature engineering for supervised learning \cite{KSWW2020}.

\subsubsection{Reinforcement Learning}
As a subcategory of ML, reinforcement learning (RL) aims to solve a variety of decision-making and control problems in a data-driven manner. Specifically, RL is able to learn in a trial-and-error way and does not require explicit human labeling or supervision of each data sample. Instead, it requires a well-defined reward function to obtain reward signals throughout the learning process. The core of RL is to learn optimal control actions in an environment to maximize the long-term cumulative reward \cite{KSWW2020}, i.e., it can be considered the model-free counterpart of model predictive control.

\subsection{Scope}

\begin{figure}[!t]
\centering
\includegraphics[width=3.4in]{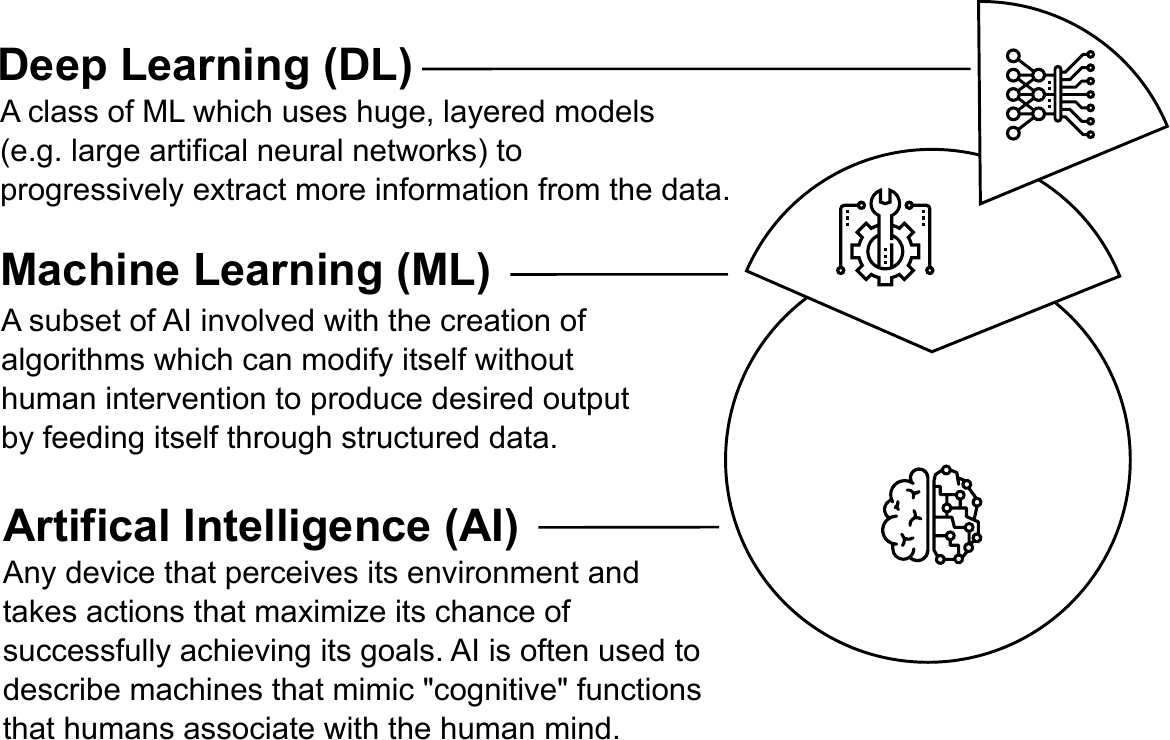}
\caption{The broader scope around ML \cite{KSWW2020}.}
\label{fig:ml_ai}
\end{figure}

The scope of this paper is to provide a comprehensive overview of the pertinent literature that applies ML techniques to electric machine drives from the 1980s to the state of the art. Despite the widespread application of classical artificial intelligence (AI) techniques in the field of electric machine drives, such as expert systems \cite{vas1999artificial}, fuzzy logic systems \cite{mir1994fuzzy, grabowski2000simple, gadoue2007genetic, suetake2010embedded, naik2015three, singh2020interval, lin2000decoupled, gadoue2009artificial, gadoue2009mras, ramesh2015type}, and evolutionary algorithms applied to tuning classical controller/estimator parameters \cite{gadoue2007genetic, demirtas2009dsp, lin2007recurrent, lin2008recurrent, hannan2018quantum, hannan2020role, hannan2018optimization}, it is in the authors' humble opinion that AI is not used here by definition, since the usual procedures are rather simple compared to the cutting-edge research within AI computer science, and resulting algorithms do not stand up to the usual definitions of ``intelligence'' that mimic ``cognitive'' functions such as perception, attention, memory or language processing \cite{colom2022human}, as illustrated in Fig. \ref{fig:ml_ai}. On the other hand, ``machine learning'' is defined as the study of (computer) algorithms that can improve automatically through experience and by the use of data, which is considered a more appropriate summary of the majority of literature included in this review paper. Therefore, ``machine learning'' will be used for the remainder of this manuscript, even if some authors might have used the exact wording of AI in the titles of their papers.

Furthermore, it is also shown in Fig. \ref{fig:ml_ai} that the general understanding of ``deep'' ML or deep learning is typically characterized by using data-driven models of (very) large depths (e.g., many ANN layers with millions of tunable parameters) such that the manual feature engineering is automatically handled within the ML pipeline (e.g., using pooling). In contrast, classical (shallow) ML comes along with manual feature engineering in expert-driven or heuristic pre-processing steps and ML models of very limited depth. Because ``genuine'' deep ML models are dramatically computationally expensive, they will remain non-real-time capable for years or even decades with regard to embedded applications in motor drives. Therefore, the scope of deep learning will also not be covered in this paper.

%
%
%
\begin{figure}[!t]
\centering
\includegraphics[width=3.4in]{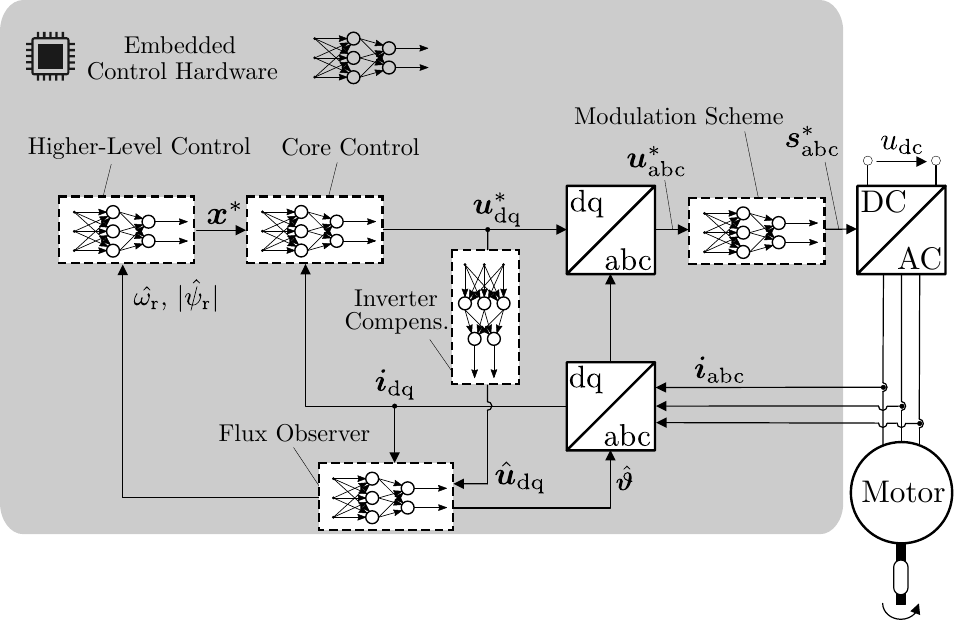}
\caption{Simplified block diagram on ML applications in a generic motor drive system. Every part of the motor drive control scheme could be ML-based, while also the entire control framework could be just one large ML model. The control depiction utilizing rotating $dq$-coordinates is only of illustrative purpose as the ML-based method is not limited to this coordinate system.}
\label{fig:ML_block_diagram}
\end{figure}
\subsection{Contribution}

The contribution of this paper is to comprehensively summarize the recent advances in applying ML-based methods to the control and monitoring of electric machine drives and to identify suitable embedded systems for deploying such ML algorithms in real time. Essentially, every part of a generic motor drive system, such as the core current/speed controller, the higher-level controller generating the optimal torque, flux, or speed commands, the flux estimator, the inverter nonideality compensation, and the modulation scheme, can be substituted by ML-based models as shown in Fig. \ref{fig:ML_block_diagram}. Additionally, the entire control framework could also be accomplished using just one large ML model. These ML-based models should normally be executed at the same task frequency as the motor control software, i.e., in the micro- to millisecond range. Classical field-oriented approaches or model-predictive control are typically used in this context. However, most ML methods do not rely on a specific control scheme.
\subsection{Outline}
The paper is organized as follows: Sections \ref{sec:IM_drives} and \ref{sec:PM_drives}
introduce specific applications of ML methods developed for induction machines and permanent magnet synchronous machines. As the inverter and sensors are important parts of any drive system, Section \ref{sec:Drive_components} discusses state-of-the-art ML techniques applied to those drive components.
In Section \ref{sec:Future_drives}, the future trend of electric machine drives enabled by state-of-the-art reinforcement learning algorithms is introduced. Section \ref{sec:Embedded} presents an in-depth comparative study on the potential embedded platforms to host such ML applications in electric machine drives for optimal cost and performance.
%
%
%

Due to the widespread popularity of ML and the abundant resources regarding its fundamentals, this review paper assumes that readers have a sufficient understanding of the basic concepts of ML, as outlined in \cite{vas1999artificial, bose2007neural}, so it can be more pivoted to introducing their successful and diverse applications in electric machine drives.

%

\section{ML-Based Induction Machine Drives}\label{sec:IM_drives}
\subsection{Classical PI/PID Controllers Replaced by ML-Based Controllers}

While the conventional proportional-integral (PI) or proportional–integral–derivative (PID)-type controllers are widely used in the industry due to their simple control structure, ease of design, and inexpensive computation cost \cite{aastrom2001future, gadoue2007genetic}, they often cannot provide perfect control performance if the controlled plant is nonlinear and uncertain \cite{wai2003wavelet}. Moreover, when using a standard field-oriented control (FOC) framework a certain voltage margin is required to ensure proper decoupling which reduces the achievable power density of a given drive \cite{7409195}.
Therefore, many ML-based controllers are designed and implemented as alternatives to the conventional PI/PID controllers to identify and adaptively control induction machines.

The idea of using artificial neural networks (ANN) to control inverter drives was first proposed in \cite{harashima1989application, buhl1991design}, where ANNs are trained offline to mimic the behavior of hysteresis current controllers to generate desired switching patterns. It is found that such ANN controllers can deliver similar performance to the original hysteresis controllers, plus certain advantages such as enhanced fault tolerance to the lack of one-phase current error input \cite{harashima1989application}. It is also worthwhile to mention that the scope of fault tolerance is only narrowly defined in \cite{harashima1989application}, as ML techniques generally cannot extrapolate to unseen events and, therefore, one would need to spend a lot of resources to train such an ML-based controller for many potential fault scenarios to make it truly fault-tolerant. Additionally, these early works have not made attempts to design an ML-based controller with better dynamic performance.

The first attempt using ANN to identify the induction machine dynamics and then control its stator currents and rotor speed in an adaptive manner is presented \cite{wishart1995identification}. For both control schemes, observable forms of the electromagnetic model of the induction machine are presented, and two systems are introduced to identify the model and the change in rotor speed using ANNs. Based directly on these two identification models, two ANN controllers are trained to adaptively control the stator currents and the motor speed. It is shown in simulation that the response of the ANN-controlled system improves with time as the system learns, and during the last transient, it actually outperforms a fine-tuned vector control system. 

Besides the specific ML-based controller scheme proposed in \cite{wishart1995identification}, there are also many other variants of such controllers that offer decent dynamic performance. For example, a two-degree-of-freedom (2DOF) controller is adopted in \cite{kung1995adaptive} to regulate the rotor speed and the stator currents as an alternative to the conventional PI controller. The controller parameters are adaptively tuned in real-time using neural networks, which can offer much improved transient performance when compared with fixed-gain 2DOF controllers.  Furthermore, a robust speed controller based on the recurrent neural network is developed in \cite{ren2006robust}. Nevertheless, it pretty much follows the same control architecture by having a recurrent neural network identifier and a recurrent neural network controller. 

However, it is also reported in \cite{rubaai2000adaptive} that such a control scheme involving two distinct neural networks in charge of the system identification and the control might lead to inadequate performance in the presence of rapid load changes. Therefore, it is recommended that the two separate tasks of system identification and control be combined into a single operation enabled by a single ANN, though no comparison results are provided to justify such claims. In \cite{rubaai2001online}, the same authors further propose using five feedforward ANNs trained in parallel, instead of one, to perform such a distinct neural-network-based estimation and control scheme. A rigorous comparative study of neural network controllers against PI controllers is presented in \cite{fu2015novel}, where both PI controllers for the $d$- and $q$-axis current can be replaced with ANN controllers as shown in Fig. \ref{fig:ML_block_diagram}. The simulation results demonstrate that the ANN-based controllers can provide better current tracking ability than PI controllers with fewer oscillations and low harmonics, and they are also less vulnerable to detuning effects caused by the variation of rotor time constant during high temperatures or at deeply saturated conditions. 

\subsection{ML-Based State Estimation for the Field-Oriented Control of Induction Machines}
For the rotor field-oriented control, it is necessary to know the instantaneous magnitude $|\hat{\psi}_{\mathrm{r}}|$ and position $\hat{\theta}$ of the rotor flux. In the direct FOC scheme, both of them need to be directly estimated based on the IM voltage model, the IM current model, or the ML-based flux observer shown in Fig.~\ref{fig:ML_block_diagram}.

Specifically, the IM voltage model in the stationary reference frame can be written as
\begin{equation}
\begin{aligned}
&v_{\upalpha \mathrm{s}}=R_{\mathrm{s}} \cdot i_{ \mathrm{\upalpha s}}+\sigma L_{\mathrm{s}} \frac{\mathrm{d} i_{ \mathrm{\upalpha s}}}{\mathrm{d} t}+\frac{L_{\mathrm{m}}}{L_{\mathrm{r}}} \cdot \frac{\mathrm{d} \psi_{ \mathrm{\upalpha r}}}{\mathrm{d} t}, \\
&v_{ \mathrm{\upbeta s}}=R_{\mathrm{s}} \cdot i_{ \mathrm{\upbeta s}}+\sigma L_{\mathrm{s}} \frac{\mathrm{d} i_{ \mathrm{\upbeta s}}}{d t}+\frac{L_{\mathrm{m}}}{L_{\mathrm{r}}} \cdot \frac{\mathrm{d} \psi_{\mathrm{\upbeta r}}}{\mathrm{d} t},
\end{aligned}
\label{eqn:IM_voltage_model}
\end{equation}
where $v_{\mathrm{\upalpha s}}, v_{\mathrm{\upbeta s}}$ are the stator voltage components, $i_{\mathrm{\upalpha s}}, i_{\mathrm{\upbeta s}}$ are the stator current components, and $\psi_{\upalpha s}, \psi_{\upbeta s}$ are the reference rotor flux linkage components all expressed in the stationary reference frame. $L_{\mathrm{m}}$ is the machine mutual inductance, $R_{\mathrm{s}}$ is the stator resistance, $L_{\mathrm{s}}$ is the stator self-inductance, $L_{\mathrm{r}}$ is the rotor self-inductance, and $\sigma$ is the leakage coefficient given by $\sigma=1-L_{\mathrm{m}}^{2} /\left(L_{\mathrm{s}} L_{\mathrm{r}}\right)$.

Additionally, the IM current model in the stationary reference frame can be written as
\begin{equation}
\begin{aligned}
\frac{\mathrm{d} \hat{\psi}_{\upalpha \mathrm{r}}}{\mathrm{d} t} & = \frac{L_{\mathrm{m}}}{T_{\mathrm{r}}} i_{\upalpha \mathrm{s}} - \frac{1}{T_{\mathrm{r}}} \hat{\psi}_{\upalpha \mathrm{r}} - {\omega}_{\mathrm{r}} \hat{\psi}_{\upalpha \mathrm{r}}, \\
\frac{\mathrm{d} \hat{\psi}_{\upbeta \mathrm{r}}}{\mathrm{d} t} & = \frac{L_{\mathrm{m}}}{T_{\mathrm{r}}} i_{\upbeta \mathrm{s}} - \frac{1}{T_{\mathrm{r}}} \hat{\psi}_{\upbeta \mathrm{r}} + {\omega}_{\mathrm{r}} \hat{\psi}_{\upbeta \mathrm{r}},
\end{aligned}
\label{eqn:IM_current_model}
\end{equation}
where $T_{\mathrm{r}}$ is the rotor time constant, ${\omega}_{\mathrm{r}}$ is the measured or estimated rotor speed, $\hat{\psi}_{\upalpha \mathrm{r}}$ and $\hat{\psi}_{\upbeta \mathrm{r}}$ are the estimated rotor flux linkage components in the stationary reference frame. 

It is also well-understood that the accuracy of the voltage model suffers at low frequencies due to the presence of ideal integration, which is susceptible to the measured input voltage bias and uncertainties on the stator resistance. However, its performance at high speeds is much more reliable as the effective voltage drop across the stator resistance becomes negligible when compared with the back-EMF. The current model, on the other hand, tends to have good accuracy at lower speeds due to the advantage that such ideal integration is not required. However, its dependence on the rotor time constant $T_\mathrm{r}$, which varies widely due to temperature-incurred variations of $R_\mathrm{r}$ and magnetic saturation-incurred variations of $L_\mathrm{r}$. Therefore, these two models are usually blended into a hybrid model to cover the whole frequency range \cite{bose2020power}\footnote{Frequently referred to as the Gopinath observer approach.}.

In attempts to overcome these issues, various ML-based state estimation schemes in the form of flux or speed observers are proposed for the rotor field-oriented vector control of induction machines.
\vspace{0.08in}
%
%
\subsubsection{ML-Based Flux Observers for the Rotor-Flux-Oriented Indirect Vector Control}

One of the earliest implementations of an ML-based flux estimator for the rotor field-oriented indirect vector control is presented in \cite{toh1994flux}, where a three-layer ANN with 20, 10, and 1 neurons is trained for different load torque transient response cases using the stator current $i_{\mathrm{ds}}, i_{\mathrm{qs}}$ in the synchronous reference frame. The output of the neural network is either the estimated flux magnitude $\hat{\psi}$ or a unit vector of the slip angle $\sin \theta_{\mathrm{sl}}$, which can further be used to calculate the unit vectors of the synchronous reference frame $\cos \theta_\mathrm{e}$ and $\sin \theta_\mathrm{e}$ with the measured rotor speed $\omega_\mathrm{r}$. The test results have successfully demonstrated the high accuracy attainable by the neural network flux estimator with the maximum absolute error of 0.03 p.u. and with an RMS error of 0.1\%, which validates that data-driven neural network flux estimators may be a feasible alternative to other flux estimation methods based on models derived by experts based on pre-knowledge.

At around the same time, \cite{theocharis1994neural} proposes a neural flux observer scheme consisting of two neural networks, namely the neural flux estimator and the neural stator estimator. While the neural flux estimator is trained in a similar fashion to estimate the rotor flux magnitude, the proposed neural stator estimator is able to continuously tune the rotor time constant $T_\mathrm{r}=L_\mathrm{r}/R_\mathrm{r}$ for generating an accurate slip frequency command $\omega_{\mathrm{sl}}^{*}$ in the indirect FOC of induction machines. Rather than estimating the rotor flux magnitude using ML-based methods, a neural network decoupling controller is designed in \cite{ba1997field} to generate the currents and slip commands ($i_{\mathrm{ds}}^*, i_{\mathrm{qs}}^*$, and $\omega_{\mathrm{sl}}^{*}$). Trained using the flux and torque commands ($\psi_\mathrm{r}^*$ and $T_{\mathrm{em}}^*$), the outputs of this three-layer ANN are compared with the outputs of the conventional decoupling controller, and the resulting errors are used to tune this neural network with either back-propagation or the Levenberg-Marquardt algorithm. Simulation results also demonstrate the accuracy of the proposed neural network decoupling controller as an alternative to the conventional indirect FOC decoupling controller of induction machines.

\vspace{0.08in}
\subsubsection{ML-Based Flux Observers for the Rotor-Flux-Oriented Direct Vector Control}
Contrary to the rotor-flux-oriented indirect direct vector control scheme where the unit vectors $\cos (\theta_\mathrm{e})$ and $\sin (\theta_\mathrm{e})$ are generated by estimating the slip frequency in a feed-forward manner, the unit vectors in the direct FOC scheme are directly estimated from the $d$ and $q$-axis components of the rotor flux linkage derived from the voltage model in \eqref{eqn:IM_voltage_model} or the current model in \eqref{eqn:IM_current_model}, and these models can also be completely or partially replaced by ML-based methods, as presented in \cite{simoes1995neural, ba1997field, marino1999linear, zhang2008stochastic}.

An ML-based flux estimator of feedback signals needed for the direct vector control is first implemented in \cite{simoes1995neural}, where a two-layer neural network with 20 neurons in the hidden layer is trained using the estimated stator flux ($\hat{\psi}_{\mathrm{\upalpha s}}$ and $\hat{\psi}_{\mathrm{\upbeta s}}$) by integrating the back-EMF and the measured stator currents ($i_{\mathrm{\upalpha s}}$ and $i_{\mathrm{\upbeta s}}$) transformed into the stationary reference frame, and the outputs are estimations of feedback signals including the magnitude of the rotor flux $|\hat{\psi}_{\mathrm{r}}|$, unit vectors $\cos (\theta_\mathrm{e})$ and $\sin (\theta_\mathrm{e})$, and torque $\hat{T}_{\mathrm{em}}$. Despite exhibiting certain advantages over the conventional flux estimator, such as faster execution speed, harmonic ripple immunity, and fault tolerance characteristics, the proposed neural flux estimator also brings an increased amount of fluctuation and noise in all of the estimated signals. This happens because the neural flux observer proposed in \cite{simoes1995neural} is designed as a pattern recognition system without any adaptation mechanism. To overcome this issue, \cite{marino1999linear} expands the training set and exploits information on the variation or detuning of the motor parameters obtained via simulation. Specifically, random noise within 10\% of the reference voltage is added to the stator voltage to enhance the variety of the training set in the neighborhood of the desired operating conditions. Moreover, the motor parameters are also varied within a suitably designed region in the parameter space. The implemented neural network flux observer has 4 inputs, 3 output neurons, and a single hidden layer with 20 neurons. 

Besides developing an ANN-based rotor flux estimator for the indirect FOC, \cite{ba1997field} also presents a neural stator flux estimator for the direct FOC to replace the conventional method that requires the integration of the back-EMF. With the IM drive in operation, measurements of input signals ($v_\mathrm{s}$ and $f$) and output responses ($i_\mathrm{s}$, $\psi_\mathrm{s}$, and $\omega_\mathrm{r}$) are taken. These signals, which inherently include parameter variations and saturation of the motor, are used to train an ANN to identify the inverse dynamics of the motor until the sum-squared error of the $a$ and $b$ phase stator flux ($\psi_{\mathrm{as}}$ and $\psi_{\mathrm{bs}}$) is below the desired level. Then the rotor flux can be calculated from the estimated stator flux using 
\begin{equation}
\begin{aligned}
\hat{\psi}_{\mathrm{dr}}&=\frac{L_{\mathrm{lr}}}{L_\mathrm{m}} (\hat{\psi}_{\mathrm{ds}}-\sigma L_\mathrm{s} i_{\mathrm{ds}}), \\
\hat{\psi}_{\mathrm{qr}}&=\frac{L_{\mathrm{r}}}{L_\mathrm{m}} (\hat{\psi}_{\mathrm{qs}}-\sigma L_\mathrm{s} i_{\mathrm{qs}}), \\
|\hat{\psi}_{\mathrm{r}}|&=\sqrt{(\hat{\psi}_{\mathrm{dr}})^{2}+(\hat{\psi}_{\mathrm{qr}})^{2}},\\
\end{aligned}
\label{eqn:IM_rotor_flux}
\end{equation}
where $\hat{\psi}_{\mathrm{dr}}$ and $\hat{\psi}_{\mathrm{qr}}$ are the estimated rotor flux components expressed in the rotor reference frame, and $\sigma$ is the leakage coefficient of the induction machine defined earlier. The unit vectors can thus be calculated as 
\begin{figure}[!t]
\centering
\def\svgwidth{\columnwidth}
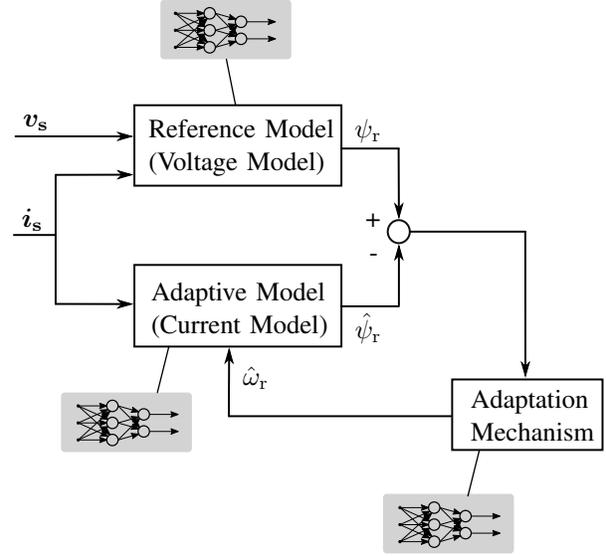
\caption{Illustration of the rotor flux-based MARS components replaced by neural networks.}
\label{fig:rotor_flux_NN_MRAS_1}
\end{figure}
%
%
%
%
%
%
\begin{equation}
\begin{aligned}
\cos \left(\theta_{\mathrm{e}}\right) &=\frac{\hat{\psi}_{\mathrm{dr}}}{|\hat{\psi}_{\mathrm{r}}|}, \\
\sin \left(\theta_{\mathrm{e}}\right) &=\frac{\hat{\psi}_{\mathrm{qr}}}{|\hat{\psi}_{\mathrm{r}}|}.
\label{eqn:flux_unit_vector}
\end{aligned}
\end{equation}

However, it should be noted that direct measurement of the stator flux used to train the neural network requires the induction motor to be modified to install flux sensors, such as Hall-effect devices and search coils, which is not appropriate for general-purpose industrial motors. Additionally, by using the model-based motor equations in  \eqref{eqn:IM_rotor_flux}, it is assumed that parameters $L_\mathrm{r}$ and $L_\mathrm{m}$ are weakly affected by saturation, which might not be the case for many highly-utilized induction machines, e.g.,  designed for the automotive industry. To overcome this issue, \cite{9432615} proposes a hybrid ML model with a structured ANN which allows estimating of both the stator flux as well as the electromagnetic machine torque thanks to introducing a priori expert knowledge on the system dynamics. Here, the stator flux is only estimated as an intermediate quantity while only a torque measurement (and not a stator flux sensor) is required to perform the data-driven training.
%

%

\subsection{ML-Based Rotor Flux Model Reference Adaptive System (MRAS) Speed Observer}

The conventional rotor flux-based model reference adaptive
system (MRAS) estimator is introduced in \cite{schauder1989adaptive}, and the structure of which is shown in Fig. \ref{fig:rotor_flux_NN_MRAS_1}. This speed observer mainly consists of two mathematical models -- the reference model and the adaptive model, as well as an adaptation mechanism to produce the estimated speed. This scheme is one of the most commonly used rotor speed estimators and many attempts have been made to improve its performance according to the literature, and it is later proven from control theories that both speed and rotor flux estimation are possible using only measurements of stator electrical quantities  \cite{vaclavek2012ac}.

The reference model is typically represented by the IM voltage model in the stationary reference frame in \eqref{eqn:IM_voltage_model}, while the adaptive model is typically represented by the IM current model in the stationary reference frame in \eqref{eqn:IM_current_model}. The presence of cross-coupling in the speed-dependent components in the adaptive model \eqref{eqn:IM_current_model} can lead to an instability issue \cite{jansen1994physically, zbede2016model}, therefore, it is common to use the rotor flux equations represented in the rotor reference frame as
\begin{equation}
\begin{aligned}
&\hat{\psi}_{\mathrm{dr}}=\frac{L_{\mathrm{m}}}{1+T_{\mathrm{r}} \cdot s} i_{\mathrm{ds}}, \\
&\hat{\psi}_{\mathrm{qr}}=\frac{L_{\mathrm{m}}}{1+T_{\mathrm{r}} \cdot s} i_{\mathrm{qs}},
\end{aligned}
\label{eqn:IM_rotor_flux_rotor_frame}
\end{equation}
where $i_{\mathrm{ds}}$ and $i_{\mathrm{qs}}$ are the stator current components, $\hat{\psi}_{\mathrm{dr}}$ and $\hat{\psi}_{\mathrm{qr}}$ are the rotor flux components all expressed in the rotor reference frame. 

The design of the adaptation mechanism is based mainly on Popov's hyperstability theory, and as a result of applying this theory, the signal of the speed tuning error $\varepsilon_\upomega$ can be written as \cite{vas1998sensorless}
\begin{equation}
\varepsilon_\upomega=\hat{\psi}_{\mathrm{\upalpha r}} \psi_{\mathrm{\upbeta r}}-\hat{\psi}_{\mathrm{\upbeta r}} \psi_{\mathrm{\upalpha r}}.
\label{eqn:IM_mras_speed_error}
\end{equation}

A PI controller is typically used to minimize this error, which in turn generates the estimated speed at its output \cite{vas1998sensorless}
\begin{equation}
\hat{\omega}_{\mathrm{r}}=\left(K_{\mathrm{p}}+\frac{K_{\mathrm{i}}}{s}\right) \varepsilon_\upomega.
\label{eqn:IM_mras_pi}
\end{equation}

Despite being a simpler and less computationally intensive method when compared with many other sensorless control methods, the main problems associated with it lie in its low-speed performance due to machine parameter sensitivity, stator voltage/current acquisition, inverter nonlinearity, and pure integration for the stator flux. Since many model-based estimation techniques rely on the back-EMF voltage, which is very small and even vanishes at zero stator frequency, these techniques will fail at or around zero speed \cite{vas1998sensorless}. To overcome these issues, various ML-based rotor flux MRAS speed observers are proposed in the literature \cite{ben1993identification, ben1995motor, ben1999speed, cirrincione2004new2, cirrincione2005mras, cirrincione2012mras, gadoue2009sensorless, kim2001speed, gadoue2009mras, ramesh2015type}. 

\vspace{0.08in}
\subsubsection{Adaptive Current Model Replaced by ML-Based Flux Observers}
Some of the earliest attempts in designing ML-based rotor flux MRAS speed observers are presented in \cite{ben1993identification, ben1995motor, ben1999speed}, where a two-layer ANN is proposed to replace the conventional adaptive current model described in \eqref{eqn:IM_current_model}. The estimated rotor flux from the ANN is compared with its target value from the reference voltage model, and the total error between the target and the estimated rotor flux is then back-propagated to adjust the weights of the neural network, after which the ANN's output will coincide with the desired value. Instead of using the classical adaptation mechanism for speed estimation as outlined in \eqref{eqn:IM_mras_speed_error} and \eqref{eqn:IM_mras_pi}, the estimated speed is represented as one of the ANN weights updated online using a backpropagation algorithm. 

Further enhancements of this scheme are presented in \cite{cirrincione2004new2} and \cite{cirrincione2005mras}, where an adaptive linear neural network is employed to represent the adaptive current model. Additionally, this ANN is tuned using the sampled stator currents and the rotor flux-linkage components coming from the model-based reference voltage model, indicating that such an adaptive ANN model is used in prediction mode rather than in simulation mode found in \cite{ben1993identification, ben1995motor, ben1999speed}. Both the recursive and the ordinary least square algorithms are used to train the ANN online to obtain the rotor speed information. When compared with the nonlinear back-propagation algorithm used in \cite{ben1993identification, ben1995motor, ben1999speed}, the proposed linear neural MRAS observer achieves better behavior in zero-speed operation at no load, as well as lower complexity and computational burden. A similar approach is also proposed in \cite{cirrincione2012mras} for the linear induction motor drive.  
\vspace{0.08in}
\subsubsection{Reference Voltage Model Replaced by ML-Based Flux Observers}

Despite the success and improvement of ANN-based flux observers replacing the conventional adaptive current model in the MRAS sensorless control algorithm, there are still problems with the IM drive's performance down to zero speed. For example, it is reported in \cite{ben1999speed} that the speed estimation performance is only acceptable when ``the operating frequency is bigger or equal to \SI{2}{\Hz}, or else fluctuations will exist in the speed estimation that ``may lead to the halting of the system.'' It is further revealed in \cite{cirrincione2005mras} that the maximum instantaneous speed estimation error at zero speed is above \SI{10}{\radian\per\second} with its adaptive current model replaced by an ANN, despite the fact that such error is as high as \SI{20}{1/\second} using the approach proposed in \cite{ben1999speed}. 

To improve the sensorless drive performance at low and zero speeds, \cite{gadoue2009sensorless} proposes a new MRAS scheme that employs an ANN flux observer to entirely replace the conventional reference voltage model, rather than the adaptive current model as described in the earlier methods. This method tends to work better at low and zero speeds as when compared with a voltage model-based flux observer, an ANN does not employ pure integration and is less sensitive to motor parameter variations. As illustrated in Fig. \ref{fig:rotor_flux_NN_MRAS_1}, a multilayer feedforward ANN that estimates the rotor flux from present and past samples of the terminal voltages and currents is used to replace the reference voltage model. The experimental results show a significantly improved low and zero-speed performance at no load versus the conventional MRAS approach. 
It is further revealed for a zero speed and \SI{20}{\%} load case, the speed estimation error at steady state is as low as \SI{7}{1/\min}, which is much lower than the method replacing the adaptive current model with ML-based flux observers.

%
\vspace{0.08in}
\subsubsection{Adaptation Mechanism Replaced by ML-Based Speed Estimators}
The performance deficiency of the conventional MRAS approach at low speeds due to pure integration and machine parameter variations can also be mitigated by replacing the fixed-gain PI controller used in the adaptation mechanism with ML-based control schemes \cite{kim2001speed, gadoue2009mras, ramesh2015type}. For example, a two-layer ANN is employed in \cite{kim2001speed} to replace such PI controllers, and the error between rotor flux estimations from the conventional reference voltage model and from the adaptive current model is back-propagated to the ANN to perform online training. The experimental results demonstrate satisfactory speed estimation with less than \SI{1}{\%} relative error when the induction machine is operating down to \SI{10}{1/\min}. 
%

%


\begin{table*}[!t]
    \caption{ML Applications in Induction Machine Drives.}
    \centering
\begin{tabular}{ll}
\hline
\rowcolor[HTML]{C0C0C0} 
\textbf{Applications}                   & \textbf{References}                      \\ 
\hline
\rowcolor[HTML]{EFEFEF} 
\textbf{ML-Based Control}           &        \\
\hline
Mimicking hysteresis current controllers to generate desired switching patterns  &  \cite{harashima1989application, buhl1991design}       \\
Replacing classical PI/PID current and speed controllers   & \cite{wishart1995identification, kung1995adaptive, ren2006robust, rubaai2000adaptive, rubaai2001online, fu2015novel}        \\
Generating the optimal flux command     & \cite{abdin2003efficiency, pryymak2006neural, ebrahim2010ann}                               \\
Achieving robust controller response against load disturbances   & \cite{huang1999robust, sheu1999self}                              \\
Implementing inverse optimal control     & \cite{quintero2018neural}                             \\
\hline
\rowcolor[HTML]{EFEFEF} 
\textbf{ML-Based State Estimation}           &        \\
\hline
Functioning as flux observers for the rotor-flux-oriented indirect vector control  &  \cite{toh1994flux, theocharis1994neural, ba1997field}       \\
Functioning as flux observers for the rotor-flux-oriented direct vector control  & \cite{simoes1995neural, ba1997field, marino1999linear, zhang2008stochastic, 9432615}        \\
Replacing the MRAS adaptive current model with ML-based speed observers  & \cite{ben1999speed, cirrincione2004new2, cirrincione2005mras, cirrincione2012mras}        \\
Replacing the MRAS reference voltage model with ML-based speed observers   & \cite{gadoue2009sensorless}  \\
Replacing the MRAS adaptation mechanism model with ML-based speed estimators  & \cite{kim2001speed, gadoue2009mras, ramesh2015type}        \\
Formulating a current error-based MRAS speed observer      & \cite{gadoue2008neural, orlowska2009adaptive}                                \\
Developing full-order and reduced-order speed observers    & \cite{cirrincione2006adaptive, cirrincione2007sensorless, accetta2013neural}                                  \\
Correcting the estimated rotor speed from sensorless nonlinear control & \cite{abu2003speed, wlas2005artificial}                         \\ 
\hline
\rowcolor[HTML]{EFEFEF} 
\textbf{ML-Based Signal Processing}           &        \\
\hline
Constituting a cascaded low-pass filter to obtain more accurate stator flux vectors      & \cite{da1999recurrent, pinto2001stator}      \\
Training a neural notch filter to estimate the rotor flux at low
speeds   & \cite{cirrincione2004new1}        \\
Introducing delayless finite impulse response and infinite impulse response filters      & \cite{zhao2004neural}      \\
\hline
\rowcolor[HTML]{EFEFEF} 
\textbf{ML-Based PWM Synthesis}           &        \\
\hline
Synthesizing space vector PWM      & \cite{bakhshai1996combined, pinto2000neural, pinto2001stator, mondal2002neural}                               \\
\hline
\rowcolor[HTML]{EFEFEF} 
\textbf{ML-Based Parameter and Model Identification}           &        \\
\hline
Learning the nonlinear machine saliency with respect to its load and flux levels  & \cite{wolbank2004combination, wolbank2007comparison}        \\
Compensating for saturation-induced saliencies in signal injection-based sensorless control  & \cite{garcia2007saliency, garcia2007automatic}        \\
Performing online identification and parameter estimation    & \cite{karanayil2005stator, karanayil2007online, wlas2008neural, bechouche2011novel, fan2014rotor}                           \\

\bottomrule
\end{tabular}
\label{tab:IM_applications}
\end{table*}
\subsection{ML-Based Parameter and Model Identification of Induction Machine Drives}
\subsubsection{ML-Based Saliency Tracking for the Sensorless Control of Induction Machines}
ML models can be used to learn the nonlinear dependencies of the machine saliency with respect to its load and flux levels \cite{wolbank2004combination}, which is crucial for reducing errors in the estimated rotor angle in IM drives with signal injection-based sensorless control. Different neural network types and learning methods are implemented and their performances are compared in \cite{wolbank2007comparison}. The results demonstrate that for the specific self-commissioning problem on an induction machine with closed rotor slots, the multi-layer perception network shows the best performance followed by the functional link neural network, whereas the time-delayed neural network is only applicable using an extensive amount of training data.

Similarly, a physical model-based neural network, also referred to as the structured neural network, is employed to compensate for such saturation-induced saliencies \cite{garcia2007saliency} and to perform automatic self-commissioning \cite{garcia2007automatic}. Originally proposed in \cite{seidl_thesis}, structured neural networks have their interconnections between neurons determined by the physical model, and their neuron basis functions are selected based on physical representations. Therefore, a structured neural network uses sinusoidal and cosinusoidal functions as its activation functions with physical meaning, versus a ``classical random (unstructured) feedforward neural network'' that uses generic activation functions (such as a sigmoid function). This structured neural network is also claimed to have a significantly reduced training time with a simpler structure than traditional neural networks. The experimental results in \cite{garcia2007saliency} demonstrate that the estimated rotor position error using such a structured neural network is roughly in line with those reported in \cite{wolbank2004combination} and \cite{briz2005rotor}. It is further reported in \cite{garcia2007automatic} that this technique has the advantages of reducing commissioning time and automating the process versus traditional methods such as look-up tables. 

Despite the fact that sinusoidal activation functions are not commonly used and do not fall into the generally applicable nonlinearities such as ReLU or sigmoid, they are shown to perform reasonably well on a couple of low-frequency, real-world datasets \cite{parascandolo2016taming}. In fact, sin/cos transformations are commonly used when learning in time-series cyclical data, such as the machine saliency discussed in this subsection. It is also worthwhile to mention that besides induction machines, this ML-based saliency tracking technique can also be extended to other machines, including permanent magnet machines and synchronous reluctance machines.

\subsubsection{ML-Based Online Parameter Estimation of Induction Machines}
Many supervised ML models have also been used to perform online parameter estimation to enable more reliable, robust, and high-performance IM drives \cite{karanayil2005stator, karanayil2007online, wlas2008neural, bechouche2011novel, fan2014rotor}. The performance of IM drives, especially those controlled by using indirect FOC, is inherently sensitive to the accuracy of the rotor time constant $T_{\mathrm{r}}$ used to estimate the slip frequency $\omega_{\mathrm{sl}}$. It is reported that the rotor resistance $R_{\mathrm{r}}$ may vary up to 100\% in certain applications over the entire range of operation due to rotor heating \cite{karanayil2005stator}, thus leading to compromised dynamic drive response if it is not estimated in a real-time manner.

To address the aforementioned issues, \cite{karanayil2005stator} and \cite{karanayil2007online} have proposed an online rotor resistance estimator using a simple two-layer ANN trained by minimizing the error between the rotor flux linkages based on an IM analytical voltage model and the output of this ANN. Since the analytical voltage model also requires the knowledge of stator resistance $R_{\mathrm{s}}$ that may also vary up to 50\% during operation, another online estimator for $R_{\mathrm{s}}$ has been added using either a fuzzy nonlinear mapping \cite{karanayil2005stator} or another ANN \cite{karanayil2007online}. The proposed ANN-based rotor resistance estimator was deployed onto a dSPACE DS1104 controller board and was executed in \SI{1}{kHz}, and satisfactory speed estimation can be obtained by using the proposed rotor resistance estimator. In \cite{fan2014rotor}, an online rotor resistance identification method is developed based on an Elman network, which is typically a three-layer network with the addition of a set of context units connected to the middle (hidden) layer fixed with a weight of one. These fixed back-connections could save a copy of the previous values of the hidden unit, making itself capable of adapting to time-varying characteristics and reflecting the dynamic characteristics of a system. The experimental results show that the proposed method is able to enhance the dynamic response of speed regulation of an IM drive.

Besides the rotor resistance $R_{\mathrm{r}}$, it is also essential to perform online parameter estimation of the mutual inductance $L_{\mathrm{m}}$ at any operating condition to achieve optimal drive performance. This parameter typically decreases with the saturation of the magnetic path, reflecting an inverse relationship that could be highly nonlinear. Two ANNs taking the form of a feedforward MLP and a recurrent network similar to an Elman network are proposed in \cite{wlas2008neural} to estimate the mutual inductance. Although both networks were trained using the same dataset, the simulation results revealed that the recurrent network maintains a filtering action that is advantageous during the oscillation of input data, while the feedforward network shows a smaller error between the value developed in the network and the value from the motor model. Therefore, the feedforward ANN is selected for experimental validation and it is shown that the accuracy of the speed estimation has been significantly improved, which further enhances the overall controller performance. Compared with the conventional feedforward ANNs to be trained online, the adaptive linear neural (ADALINE) network has a simpler structure with only a single-layer network, and its weights are can be interpreted physically. In \cite{bechouche2011novel}, the IM model is approximated by two first-order subsystems with appropriate assumptions at the low and the high frequency, which can be readily used to design two ANALINE networks to identify an IM's electrical parameters at standstill. After online training, the stator resistance $R_{\mathrm{s}}$ and the stator cyclic inductance $L_{\mathrm{r}}$ are identified via the low-frequency ADALINE network, while the rotor time-constant $R_{\mathrm{s}}$ and the leakage factor $\sigma$ are identified via the high-frequency network. Apart from the above papers introduced in detail, readers are also referred to a comprehensive review paper on performing online identification and parameter estimation in IM drives for more details \cite{gutierrez2013review}.

\subsection{Summary}
A summary of the aforementioned literature on ML models applied to induction machine drives is presented in TABLE \ref{tab:IM_applications}.

\section{Machine Learning-Based Permanent Magnet Synchronous Machine Drives}\label{sec:PM_drives}

\subsection{ML-Based Controllers for PMSM Drives}
A satisfactory current or speed controller should enable a PM machine drive to follow any reference signal taking into account the effects of load impact, temperature, saturation, and parameter variations. However, as presented in the earlier analysis in Section \ref{sec:IM_drives}-A, conventional PI and PID controllers can lack the structural ability to achieve these objectives under challenging real-world conditions Therefore, similar to those applied to IM drives, ML-based controllers are also proposed in the literature to improve the dynamic response of PM machine drives \cite{rahman1998line, yi2003implementation, guo2011model, lucas2004introducing, daryabeigi2009interior}. 

For example, ANNs are also implemented as speed controllers in PMSM drives based on motor dynamics and nonlinear load characteristics \cite{rahman1998line, yi2003implementation, guo2011model}. In \cite{rahman1998line} and \cite{yi2003implementation}, an ANN speed controller is developed to generate the $q$-axis current reference $i^{*}_\mathrm{q}(n)$, and the input of which are selected as the speed of the motor at the present and previous two sample intervals $\left[\omega_\mathrm{r}(n), \omega_\mathrm{r}(n-1), \omega_\mathrm{r}(n-2)\right]$ in addition to the previous sample of the $q$-axis current reference $i^{*}_\mathrm{q}(n-1)$. The ANN speed controller can be integrated into the vector control scheme of the PMSM drive with the initial weights and biases obtained through the offline training using simulation data. Then these weights and biases are updated online when an error between the actual output and the target of the ANN exceeds a preset value. The robustness of the proposed ANN scheme is validated using experiments against parameter variations \cite{rahman1998line} and load disturbances \cite{yi2003implementation} in real time. In addition, an ANN-based speed controller consisting of a radial basis function network is proposed in \cite{guo2011model} and the network is trained online to adapt to system uncertainties. The error between the reference speed and the measured speed is fed into the proposed ANN-based controller and its weights and biases are trained online. The experimental results with load disturbances demonstrate that the proposed ANN speed controller is able to regulate the motor speed in a more stable manner and with fewer transients when compared with the conventional PI controller. Furthermore, a brain emotional learning-based intelligent controller is further proposed in \cite{lucas2004introducing} and \cite{daryabeigi2009interior} to control the motor speed with very fast response and robustness with respect to disturbances and manufacturing imperfections.

Other relevant literature on this topic includes the hardware/software controller designs using fuzzy ANNs for brushless DC motor drives \cite{rubaai2000continually, rubaai2002development, rubaai2008design, rubaai2008dsp, rubaai2011ekf, rubaai2015hardware}, achieving robust controller response \cite{wai2001total, lin2008robust}, as well as formulating sliding mode \cite{lin2010fpga} and adaptive control schemes \cite{lin2008adaptive, abuhasel2019adaptive, el2019robust} for PM motor drives. 
\subsection{ML-Based State Estimation for PMSM Drives}
\subsubsection{ML-Based State Estimation for the Sensorless Control of PMSM Drives}
A number of classical state estimation methods have been developed to achieve the sensorless control of PMSM drives, such as state observers, Kalman filters, disturbance observers, MRAS observers, sliding-mode observers, high-frequency signal injection \cite{linke2002sensorless, linke2003sensorless, holtz2008acquisition}, etc. However, these techniques usually suffer from DC drift due to motor parameter variations and the influence of inverter nonlinearities \cite{xu2018review}. To overcome these issues, a wide variety of ML-based methods are implemented to improve the existing sensorless control schemes.

Similar to the MRAS method for induction machines in Section \ref{sec:IM_drives}-C, the MRAS for PM machines also needs an adaptation mechanism to provide accurate speed and position estimations. However, the conventional adaptation mechanism is mostly linear, making it challenging to account for the effects of torque constant and stator resistance variations on the rotor speed and position estimations. Therefore, a two-layer ANN is implemented in \cite{elbuluk2002neural} as the nonlinear adaptation mechanism as shown in Fig. \ref{fig:rotor_flux_NN_MRAS_1}, and the experimental results demonstrate that the proposed method is able to track these varying parameters at different speeds with consistent performance. 

ML methods have also been extensively applied to improve the popular sliding-mode observer (SMO) designed using the extended EMF model of PM machines \cite{lin2014sensorless, zhang2015adaline, makni2016rotor, zine2017hybrid, zine2017interests, wang2017sliding}. The voltage equation of the PM machine in the stationary reference frame can be expressed as \cite{chen2003extended}
\begin{equation}
\begin{bmatrix}
v_{\upalpha \mathrm{s}} \\
v_{\upbeta \mathrm{s}}
\end{bmatrix}
=
\begin{bmatrix}
R_{\mathrm{s}}+p L_{\mathrm{d}} & \omega_\mathrm{e}\left(L_{\mathrm{d}}-L_{\mathrm{q}}\right) \\
-\omega_\mathrm{e}\left(L_{\mathrm{d}}-L_{\mathrm{q}}\right) & R_{\mathrm{s}} + p L_{\mathrm{q}}
\end{bmatrix}
\begin{bmatrix}
i_{\upalpha \mathrm{s}} \\
i_{\upbeta \mathrm{s}}
\end{bmatrix}
+
\begin{bmatrix}
e_{\upalpha s} \\
e_{\upbeta s}
\end{bmatrix},
\label{eqn:PM_voltage_equation}
\end{equation}
where $v_{\mathrm{\upalpha s}}, v_{\mathrm{\upbeta s}}$ are the stator voltage components, $i_{\mathrm{\upalpha s}}, i_{\mathrm{\upbeta s}}$ are the stator current components, and $e_{\upalpha s}, e_{\upbeta s}$ are the extended EMF components all expressed in the stationary reference frame. $L_{\mathrm{d}}$ and $L_{\mathrm{d}}$ are the inductance of the $d$- and $q$-axis, $R_{\mathrm{s}}$ is the stator resistance, $\omega_\mathrm{e}$ is the rotor electrical speed, and $p$ is the differential operator, respectively.

The extended EMF is defined as
\begin{equation}
\begin{bmatrix}
e_\upalpha \\
e_\upbeta
\end{bmatrix}
=
\left[\left(L_{\mathrm{d}}-L_{\mathrm{q}}\right)\left(\omega_\mathrm{e} i_{\mathrm{d}}-p i_{\mathrm{q}}\right)+\omega_{\mathrm{e}} \psi_\mathrm{f}\right]
\begin{bmatrix}
-\sin(\theta_\mathrm{e}) \\
\cos(\theta_\mathrm{e})
\end{bmatrix},
\label{eqn:PM_extended_emf}
\end{equation}
where $\theta_\mathrm{e}$ is the rotor position and $\psi_\mathrm{f}$ is the permanent magnet flux linkage.

An SMO can then be designed based on the extended EMF model of the PM machine in \eqref{eqn:PM_voltage_equation} to extract the rotor spatial information contained in \eqref{eqn:PM_extended_emf} as %
\begin{equation}
\dot{\hat{i}}_\mathrm{s}=\boldsymbol{A} \hat{\boldsymbol{i}}_\mathrm{s}+\boldsymbol{B}\left(\boldsymbol{u}_\mathrm{s}-\boldsymbol{z}\right),
\end{equation}
where 
\begin{equation}
\begin{array}{l}
\boldsymbol{A}=
\begin{bmatrix}
-R_{\mathrm{s}} / L_{\mathrm{d}} & \hat{\omega}_\mathrm{e}\left(L_{\mathrm{q}}-L_{\mathrm{d}}\right) / L_{\mathrm{d}} \\
-\hat{\omega}_\mathrm{e}\left(L_{\mathrm{q}}-L_{\mathrm{d}}\right) / L_{\mathrm{d}} & -R_{\mathrm{s}} / L_{\mathrm{d}}
\end{bmatrix}
 \\
 \\
\boldsymbol{B}=
\begin{bmatrix}
1 / L_{\mathrm{d}} & 0 \\
0 & 1 / L_{\mathrm{d}}
\end{bmatrix},
~\hat{\boldsymbol{i}}_\mathrm{s}=
\begin{bmatrix}
\hat{i}_{\upalpha s} \\
\hat{i}_{\upbeta s}
\end{bmatrix}
~\boldsymbol{u}_\mathrm{s}=
\begin{bmatrix}
u_{\upalpha s} \\
u_{\upbeta s}
\end{bmatrix},
\end{array}
\end{equation}
and $\hat{\omega}_\mathrm{e}$ is the estimated electrical speed. The sliding-mode control function $\boldsymbol{z}$ contains the useful rotor position information and is defined as $\boldsymbol{z} = g \cdot F(\hat{\boldsymbol{i}}_\mathrm{s} - \boldsymbol{i}_\mathrm{s})$, where $g$ is the gain of the control function and $F(\hat{\boldsymbol{i}}_\mathrm{s} - \boldsymbol{i}_\mathrm{s})$ can be a signum, saturation, or sigmoid function \cite{wang2012quadrature}. With $F(\hat{\boldsymbol{i}}_\mathrm{s} - \boldsymbol{i}_\mathrm{s})$ being selected as a saturation function and the gain of the control function $g$ being greater than the maximum value of the extended EMF, namely $g>|e|_{\mathrm{max}}$, the observer can be asymptotically stable and the state can converge to $S = \boldsymbol{i}_\mathrm{s} - \boldsymbol{i}_\mathrm{s} = \mathbf{0}$ in a finite time. Therefore, the relation between the estimated EMF and the control function becomes \cite{zhang2015adaline}
\begin{equation}
\hat{\boldsymbol{e}}_\mathrm{s} = \boldsymbol{z}.
\end{equation}

Conventionally, the position estimate can be calculated directly from EMF estimates through an arc-tangent function as
\begin{equation}
\hat{\theta}_{\mathrm{e}}=-\tan ^{-1}\left(\hat{e}_{\upalpha s} / \hat{e}_{\upbeta s}\right).
\end{equation}

However, the presence of noise signals may adversely affect the accurate estimation of rotor position. This is especially the case when using the arc-tangent function during the zero-crossing of EMF signals. To improve the position estimation for industrial applications, a software phase-lock-loop (PLL) is typically used to obtain rotor position from the estimated EMF information. Moreover, normalization of the EMF is often necessary for the position observer due to the magnitude of the EMF varying at different velocities. In this way, the equivalent position error signal of the EMF model can be obtained as
\begin{equation}
\varepsilon_\mathrm{f}=\frac{1}{\sqrt{e_{\upalpha s}^2+e_{\upbeta s}^2}}\left[-e_{\upalpha s} \cos \left(\hat{\theta}_{\mathrm{e}}\right)-e_{\upbeta s} \sin \left(\hat{\theta}_{\mathrm{e}}\right)\right].
\end{equation}
\begin{figure}[!t]
\centering
\includegraphics[width=3.4in]{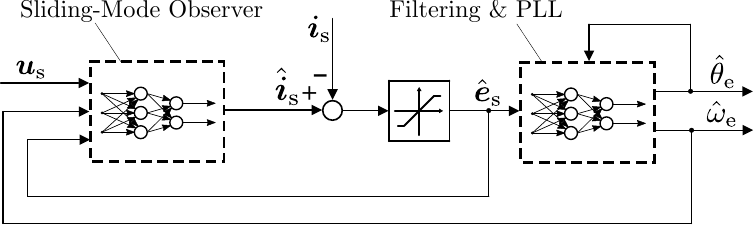}
\caption{Simplified block diagram on ML serving as different components of the SMO-based position observer based on the extended EMF model of PM machines.}
\label{fig:ML_SMO}
\end{figure}

Therefore, the position observer based on the software PLL can be expressed as
\begin{equation}
\hat{\theta}_{\mathrm{e}}=\left(1/s\right)\left(k_\mathrm{i} / s+k_\mathrm{p}\right) \varepsilon_\mathrm{f},
\end{equation}
where $k_\mathrm{p}$ and $k_\mathrm{i}$ are the proportional and integral gains of the PI controller commonly used for the software PLL.

Fig. \ref{fig:ML_SMO} illustrates implementations of ML models replacing different subsystems of the sensorless position observer based on the extended EMF model of PM machines. In the pertinent literature, \cite{lin2014sensorless} proposed a five-layer wavelet fuzzy neural network (WFNN) to replace the conventional PI controller in the PLL. To obtain good control performance in transient states and deal with the uncertainties of PM machines, both the rotor angle estimation error and its derivative are selected as inputs to the network. In order to train the WFNN effectively and guarantee the convergence of the rotor angle estimation error, the varied learning rates are derived based on the analysis of a discrete-type Lyapunov function. When compared with the PID-based PLL, the proposed WFNN-based PLL is able to reduce the average rotor position estimation error from \SI{4.06}{\degree} to \SI{2.22}{\degree} based on the experimental results. Due to the influence of inverter non-idealities and flux spatial harmonics, the $(6k\pm 1)$ order harmonics are typically present in EMF estimates, resulting in the $(6k\pm 1)$ harmonic ripples in the estimated rotor position and speed and compromise their accuracy. To mitigate this issue, \cite{zhang2015adaline} designed a multi-input, single-output, and single-layer ADALINE network to track and compensate for these $(6k\pm 1)$ order harmonics in the estimated EMF signals before they are fed into the PLL. By continuously updating the filter weights online, the experimental results demonstrated that this ADALINE-based filter is able to effectively suppress the $(6k\pm 1)$ harmonic ripples in the estimated rotor position and reduce its maximum position error from \SI{8.3}{\degree} to \SI{2.2}{\degree}.
Finally, \cite{Brosch2023sensorless} proposes an entirely data-driven sensorless PMSM torque control scheme which does not require any offline training. The proposed sensorless control algorithm can commission itself in a fully automated fashion, i.e., pre-knowledge regarding a specific drive system is not required. 

To design an EMF-based observer independent of any machine parameters, \cite{makni2016rotor, zine2017hybrid, zine2017interests} developed an ANN observer that is trained to map between the dataset of the inputs ($i_\upalpha, i_\upbeta, v_\upalpha, v_\upbeta$) and those of the outputs ($\sin(\theta_\mathrm{e})$, $\cos(\theta_\mathrm{e})$), followed by a PI-based PLL that tracks the rotor speed information based on the processed position error, and subsequently the rotor position by performing integration. The ANN-based observer has been tested on a 32-bit micro-controller and the inference time is around \SI{5}{\micro\second}. It is also revealed in \cite{zine2017interests} that the performance of this ANN-based position estimator is very poor at zero and low-speed regions, and there are no comparative results presented using the conventional PI-based observer. The conventional SMO, however, is also known to have compromised performance at the standstill and low-speed conditions due to the amplitude of the back-EMF being almost zero, \cite{wang2017sliding} thus integrates an ANN-based angle compensation scheme into an iterative SMO that successfully mitigates this issue. Specifically, the experimental results demonstrate that at a low speed of 100 revolutions per minute, the maximum rotor position estimation error can be reduced to \SI{4}{\degree} from around \SI{70}{\degree} using an iterative SMO without the proposed neural network, and this large error might also indicate an issue in the way that the iterative SMO is designed or implemented.
%




\subsubsection{ML-Based Temperature Estimation of PMSM Drives}
Besides estimating the states of motors that are directly related to drive controllers, temperature estimation is also a focal point for PMSM drives. This is because overheating can trigger irreversible demagnetization of permanent magnets as well as severe deterioration of other motor components, thus it is of high concern for the machine’s control strategy and will result in oversized motor and inverter designs leading to lesser device utilization and higher material cost \cite{kirchgassner2021data}.

The traditional approaches for temperature estimation are typically accomplished by using sensor-based methods or by estimating electrical parameters such as the stator winding resistance, stator inductance, permanent magnet flux linkage, etc., using state-space observers or high-frequency injection. However, precise temperature estimation, especially for predicting the latent and highly dynamic magnet temperature, still remains a challenging task, while the conventional methods still prove unfeasible in a commercial context \cite{kirchgassner2021data}. In this regard, data-driven approaches could become very competitive once proven that they could deliver highly accurate temperature estimations at low to moderate model sizes that would allow them to run in real time in embedded systems.

In the spirit of pursuing this goal, a comprehensive benchmark study has been conducted in \cite{kirchgassner2021data} that empirically evaluates the temperature estimation accuracy of permanent magnets using many classical ML models, including ordinary least squares (OLS), support vector regression, k-nearest neighbors, randomized trees, and multilayer perception (MLP) feedforward neural networks. All of these ML models have been trained using the same test bench data collected from a three-phase, \SI{52}{kW} automotive traction PMSM. This work also reveals the full potential of simpler ML models, especially linear regression and simple feed-forward neural networks with optimized hyperparameters, in terms of their regression accuracy, model size, and data demand in comparison to parameter-heavy deep neural networks, which are implemented in \cite{kirchgassner2020estimating} in the form of recurrent neural networks and temporal convolutional networks (TCN). For example, the mean squared errors of OLS and MLP are \SI{3.10}{K^2} and \SI{3.20}{K^2}, respectively, while the TCN can further reduce this error by more than 50\% to \SI{1.52}{K^2}. However, this accuracy comes at a cost of using as many as \SI{67}{k} model parameters and its inference duration is taking 115 times longer than that of the OLS. The simpler ML models of OLS and MLP, on the other hand, only have 109 and \SI{1.8}{k} model parameters with inference duration of 1.0 and 14.8, which are normalized with respect to the OLS model.

The potential of ML models can be further expanded by adding more interpretability at their design stage, thus allowing humans to capture relevant knowledge from such models concerning relationships either contained in data or learned by the model. \cite{kirchgassner2023thermal} thus introduces a novel thermal neural network (TNN) that unifies both the consolidated knowledge in the form of heat-transfer-based lumped-parameter models, and data-driven nonlinear function approximation with supervised machine learning. Experiments on the same electric motor data set show that this TNN is able to achieve accurate temperature estimation with a mean squared error of \SI{3.18}{K^2} at only 64 model parameters. A detailed review of temperature estimation methods for PMSMs, including the application of different ML models, can be found in \cite{wallscheid2021thermal}.

\subsection{ML-Based Parameter and Model Identification of PMSM Drives}
Similar to IM drives, accurate online parameter estimation and model identification are also crucial to achieving robust and high-performance PMSM drives across their entire range of operations. One of the earliest attempts at using ML-based methods for this purpose is reported in \cite{liu1998torque}, which proposes a standard, 3-layer feedforward ANN to estimate the torque constant and stator resistance of a PM motor, the values of which will be used for torque ripple minimization of a deadbeat predictive current controller. The simulation results show that the drive system is insensitive to these parameter changes after implementing this ANN-based parameter estimator and the torque ripple is reduced from 5\% to 3\%.

The simple structure and low computational demand of the ADALINE network have also been leveraged in PMSM drives for online model identification. For example, \cite{mohamed2007novel} implements a direct instantaneous torque and flux controller that requires accurate knowledge of the instantaneous electromagnetic torque, stator flux vector, and machine electrical parameters in order to accomplish a high-performance instantaneous torque control scheme. All of these quantities are estimated online using an ADALINE-based PM motor model that is trained through back-propagation by minimizing the mean squared error between the measured q-axis current and its estimated value from the ADALINE network. When compared with the conventional torque control with decoupled PI current controllers, the experimental results reveal that the torque ripple has been reduced from 8.5\% to only 0.5\% at \SI{10}{1/\min} when using the ADALINE-based motor model, and this ML-enabled drive system is able to offer fast and smooth torque response with enough robustness against disturbances and parameter variations. Similarly, an online parameter estimator based on a variable step-size ADALINE network is proposed in \cite{wang2020deadbeat} to identify the PMSM parameters such as the stator synchronous inductance $L_{\mathrm{s}}$, the stator resistance $R_{\mathrm{s}}$, and the permanent magnet flux linkage $\psi_\mathrm{f}$. The identification results of motor parameters are then substituted into the prediction model of a deadbeat predictive current controller, which eliminates the current static error caused by parameter mismatch and effectively improves the parameter robustness of the controller.%

Furthermore, \cite{brosch2020data} develops a data-driven recursive least squares estimation method for online motor parameter identification to improve the prediction accuracy of the finite-control-set model-predictive-current (FCS-MPC) control of PM drives. The PMSM model parameters can be recursively corrected with each new measurement and, therefore, the resulting FCS-MPC algorithm enabled by this data-driven method is able to outperform a baseline white-box model derived from first-order physical principles \cite{9359525}. To overcome the global forgetting of (ultra-)local models \cite{8924938}, \cite{9891791} extends the adaptive local model approach with a long-term memory to allow instant model reconfiguration to already visited operating points.  In \cite{jie2020adaptive, jie2020speed}, a novel adaptive decoupling controller is also introduced based on radial basis function neural network to estimate the uncertain and time-varying motor parameters, namely the stator resistance $R_{\mathrm{s}}$, the stator inductance of the $d-$ and $q-$ axis $L_{\mathrm{d}}$, $L_{\mathrm{q}}$, and also the permanent magnet flux linkage $\psi_\mathrm{f}$. All of these online estimations have been proven to  improve the dynamic and steady-state characteristics of the drive system.

In addition to the discussions above, readers are also referred to the review paper on ML-based online identification and parameter estimation of PM machines in \cite{rafaq2019comprehensive} for more details on this topic.
\begin{table*}[!t]
    \caption{ML Applications in Permanent Magnet Machine Drives.}
    \centering
    \begin{tabular}{lll}
    \hline
    \rowcolor[HTML]{C0C0C0} 
    Applications                   & References                      \\
    \hline
    \rowcolor[HTML]{EFEFEF} 
    \textbf{ML-Based Control}           &        \\
    \hline
    Replacing classical PI/PID current and speed controllers     & \cite{rahman1998line, yi2003implementation, guo2011model, lucas2004introducing, daryabeigi2009interior} \\
    Implementing the hardware/software designs for brushless DC motor drives    & \cite{rubaai2000continually, rubaai2002development, rubaai2008design, rubaai2008dsp, rubaai2011ekf, rubaai2015hardware}   \\
    Achieving robust controller response     & \cite{wai2001total, lin2008robust}               \\
    Implementing sliding mode control on a PM linear servo motor drive system     & \cite{lin2010fpga}      \\
    Designing adaptive controller schemes & \cite{lin2008adaptive, abuhasel2019adaptive, el2019robust}              \\
    \hline
    \rowcolor[HTML]{EFEFEF} 
    \textbf{ML-Based State Estimation}           &        \\
    \hline
    Replacing the MRAS adaptation mechanism model with ML-based speed estimators    & \cite{elbuluk2002neural}\\
    Improving different subsystems of the popular back-EMF-based observer with PLL     & \cite{lin2014sensorless, zhang2015adaline, makni2016rotor, zine2017hybrid, zine2017interests, wang2017sliding} \\
    \hline
    \rowcolor[HTML]{EFEFEF} 
    \textbf{ML-Based Parameter and Model Identification}           &        \\
    \hline
    Performing online or offline identification and parameter estimation           & \cite{liu1998torque, mohamed2007novel, wang2020deadbeat, brosch2020data, 9359525, 8924938, 9891791}   \\
    Executing adaptive decoupling control considering uncertain and time-varying parameters      & \cite{jie2020adaptive, jie2020speed}   \\
    Estimating the temperature of permanent magnets \& multiple stator components     & \cite{kirchgassner2020estimating, kirchgassner2021data, kirchgassner2023thermal}        \\
    \bottomrule
    \end{tabular}
    \label{tab:PM_applications}
\end{table*}

\subsection{Summary}
A summary of the aforementioned literature on ML models applied to induction machine drives is presented in TABLE \ref{tab:PM_applications}. The next-generation reinforcement learning-based motor control schemes, with most of the existing literature carried out on PM machines, can be found in Section \ref{sec:Future_drives}.

\section{Machine Learning Techniques Applied to Drive Inverters and Sensors}\label{sec:Drive_components}

As the inverter and sensors are also important parts of any modern drive system, this section discusses state-of-the-art ML techniques applied to those drive components, particularly in terms of modeling and compensating the inverter non-ideal characteristics \cite{stender2020data, stender2020comparison, stender2021gray, brosch2020data} and condition monitoring of sensors used to provide critical feedback signals in a motor drive system \cite{gou2018intelligent, gou2020online, xia2021learning, argawal2021sensor, dybkowski2019artificial, jankowska2022design}.
%

\subsection{ML-Based Modeling and Compensation of the Drive Inverter Non-Ideal Characteristics}
\begin{figure}[!t]
    \centering
    \subfloat[]{\includegraphics[width=3.4in]{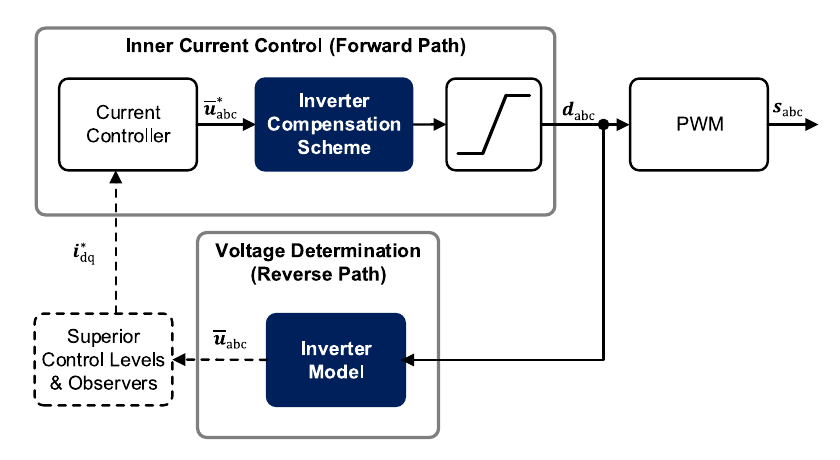}} \\
    \subfloat[]{\includegraphics[width=3.2in]{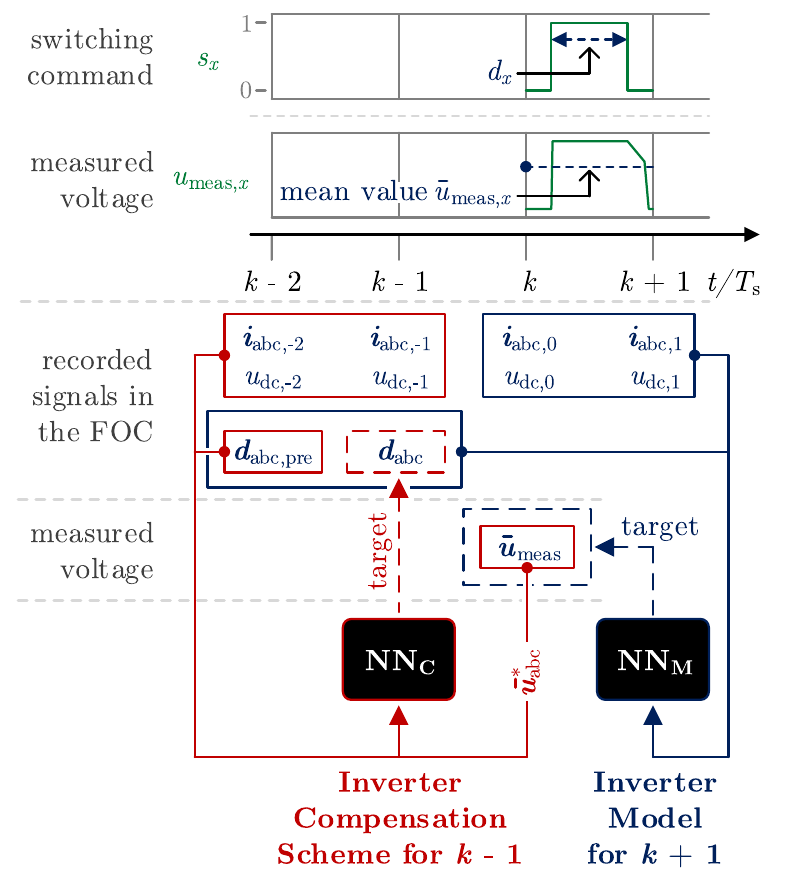}}
    \caption{(a) Schematic integration of an inverter model and inverter compensation scheme into electric machine drives, and (b) visualization of the utilized signals for the training process of the black-box inverter model with time dependence: NN$_{\mathrm{M}}$ and NN$_{\mathrm{C}}$ represent the neural networks for the inverter model \eqref{eqn:inverter_nn_model} and the inverter compensation scheme \eqref{eqn:inverter_compensation_scheme}, respectively \cite{stender2020comparison}.}
    \label{fig:inverter_model_and_compensation_scheme}
\end{figure}
In most motor drive applications, the stator phase voltages have modulated signal forms due to inverter switching and are therefore difficult to measure. While it is technically feasible to measure them directly such as using delta-sigma modulators, additional cost and integration effort have prevented the widespread implementation of phase voltage measurements in mass-produced drives \cite{wallscheid2021thermal}. Therefore, an accurate inverter model is required to estimate the phase voltages $\thickbar{\boldsymbol{u}}_{\mathrm{abc}}$ from the reference voltage information $\thickbar{\boldsymbol{u}}^*_{\mathrm{abc}}$ in the motor control algorithm, as shown in Fig. \ref{fig:inverter_model_and_compensation_scheme}(a).

However, due to various non-ideal characteristics of the drive inverter such as the interlocking time, non-ideal switching behaviors due to parasitics, signal delays, and the forward voltage drop across semiconductors and cables, an analytical white-box modeling approach requiring simulation step times in the nanosecond range is hardly feasible in a control context \cite{stender2020data}. Therefore, a black-box inverter model utilizing ML and data-driven approaches is considered favorable for this task. To train such an inverter model or compensation scheme incorporating the inverter's non-ideal characteristics, a large amount of data samples needs to be collected a priori that cover the complete operating envelope of a motor drive system. An exemplary dataset of 234,500 samples has been collected in \cite{stender2020data} on a two-level IGBT inverter and published on Kaggle \cite{stender2020kaggle}.

Based on this open dataset, a comprehensive data-driven black-box inverter model is established using a neural network \cite{stender2020comparison} to approximate the following function
\begin{equation}
\boldsymbol{f}_{\mathrm{nn}}\left(\boldsymbol{d}_{\mathrm{abc}}, \boldsymbol{d}_{\mathrm{abc}, \mathrm{pre}}, \boldsymbol{i}_{\mathrm{abc}, 0}, \boldsymbol{i}_{\mathrm{abc}, 1}, u_{\mathrm{dc}, 0}, u_{\mathrm{dc}, 1}\right)=\thickbar{\boldsymbol{u}}_{\mathrm{abc}},
\label{eqn:inverter_nn_model}
\end{equation}
which provides the set mean voltage vector $\thickbar{\boldsymbol{u}}_\mathrm{abc}$ with respect to the relevant set duty vector of the current and the previous PWM period ($\boldsymbol{d}_\mathrm{abc}$,  $\boldsymbol{d}_\mathrm{abc,pre}$), the measured phase current vectors at the beginning of the current and the next PWM period ($\boldsymbol{i}_{\mathrm{abc}, 0}$, $\boldsymbol{i}_{\mathrm{abc}, 1})$, and the measured DC-link voltage values at the beginning of the current and next PWM period ($u_{\mathrm{dc}, 0}$, $u_{\mathrm{dc}, 1}$).

Similarly, a suitable function for the inverter compensation scheme illustrated in Fig. \ref{fig:inverter_model_and_compensation_scheme}(a) can be formulated by rearranging the output variable and the input features of the inverter model
\begin{equation}
\begin{aligned}
{\boldsymbol{f}_{\mathrm{nnc}}(\thickbar{\boldsymbol{u}}_{\mathrm{abc}}^{*}, \boldsymbol{d}_{\mathrm{abc}, \mathrm{pre}}, \boldsymbol{i}_{\mathrm{abc},-1}}&, \boldsymbol{i}_{\mathrm{abc},-2} \\
& u_{\mathrm{dc},-1}, u_{\mathrm{dc},-2})=\boldsymbol{d}_{\mathrm{abc}}^{*}
\end{aligned}
\label{eqn:inverter_compensation_scheme}
\end{equation}
with the corresponding signals defined in Fig. \ref{fig:inverter_model_and_compensation_scheme}(b).
\begin{table*}[!t]
    \caption{Condition Monitoring of Motor Drive Sensors Using ML Models.}
    \begin{center}
    \resizebox{\linewidth}{!}{
    \begin{tabular}{lcccc}
    \toprule
    \rowcolor[HTML]{FFFFFF} 
    ML model            & Sensor type       & Fault type      & Inference time        & References    \\ 
    \midrule
    \rowcolor[HTML]{EFEFEF} 
    Extreme learning machine          & Current, speed, and DC-voltage sensors  &  Stuck, erratic, and offset faults       &   \SI{10}{ms}   & \cite{gou2018intelligent}           \\
    \rowcolor[HTML]{FFFFFF} 
    Random vector functional link network    &  Current sensor$^*$  &  Stuck, erratic, and offset faults    &   \SI{22}{ms} & \cite{gou2020online}                               \\
    \rowcolor[HTML]{EFEFEF} 
    Random vector functional link network    &  Speed sensor        &  Stuck, erratic, and gain faults      &  \SI{228}{ms} & \cite{xia2021learning}                         \\
    \rowcolor[HTML]{FFFFFF} 
    Decision tree, support vector machines   &  Current sensor      &  Stuck and offset faults              &  N/A          & \cite{argawal2021sensor}     \\
    \rowcolor[HTML]{EFEFEF} 
    ANN                                      &  Current sensor      &  Stuck, erratic, and gain faults      & \SI{0.8}{ms} -- \SI{2.03}{s} & \cite{dybkowski2019artificial}                               \\
    \rowcolor[HTML]{FFFFFF} 
    ANN                                      &  Current sensor      &  Stuck and gain faults                &  N/A & \cite{jankowska2022design}                               \\
    \bottomrule
    \end{tabular}}
    \end{center}
    \begin{footnotesize}
    \hspace{0.1in}$^*$Also detects IGBT faults.\\
    \end{footnotesize}
    \label{tab:drive_sensor_monitoring}
\end{table*}

For the neural network representing the ML-based inverter model, a basic feedforward network layout is chosen, while hyperparameters such as the number of hidden layers, the number of neurons, and the type of activation functions are determined by Bayesian optimization. A detailed comparative study with a gray-box inverter model combining first-order principles from physics with data-driven-based parameter identification reveals that the ML-based black-box model can precisely estimate the phase voltages per switching cycle with a root-mean-square error of less than \SI{0.65}{V} at a \SI{560}{V} of DC-link voltage level, outperforming the gray-box model that only achieves an error of less than \SI{1.1}{V} \cite{stender2020comparison}. The scope of the aforementioned gray-box inverter model has been further extended in \cite{stender2021gray} to also estimate the power losses in the motor and inverter, and the parameter of which is also obtained via a data-driven approach based on particle swarm optimization. It is envisioned that based on this precise loss modeling, an optimal motor operation strategy can be developed in response to system changes in real time during operation.


\subsection{ML-Based Condition Monitoring of Motor Drive Sensors}
Sensors are indispensable parts of any modern electric machine drive to provide accurate real-time feedback signals for enabling high-performance closed-loop controls. Specifically, current sensors and rotor speed/position sensors are typically required in most drive applications, while DC-link voltage sensors are needed for implementing advanced features such as the sensorless flux observer or online parameter estimation. However, the sensors in the drive system are prone to various faults due to aging, vibration, humidity, and surrounding interference \cite{gou2018intelligent}. The typical fault modes of the sensor can be broadly classified as stuck faults, erratic faults, gain/offset faults, drift faults, and spike faults \cite{jan2017sensor}:
\begin{enumerate}
    \item[1)] \emph{Stuck faults}: The sensor’s output gets stuck at a fixed value, and this can also be viewed as a complete failure.
    \item[2)] \emph{Erratic faults}: Variance of the sensor output significantly increases above the normal value.
    \item[3)] \emph{Offset faults}: The output of the sensor possesses a constant offset value some time after calibration.
    \item[4)] \emph{Drift/gain faults}: The output of the sensor keeps increasing or decreasing linearly from the normal state.
    \item[5)] \emph{Spike faults}: Spikes are observed in the sensor output at fixed intervals.
\end{enumerate}

Possible consequences of the above fault modes on the different sensors in the drive system can be briefly discussed as follows \cite{gou2018intelligent}. A current sensor fault can result in an imbalanced current flowing into the motor, causing overheating and fluctuation/instability in speed and torque control. A speed sensor fault affects the desired orthogonal alignment of the stator field and the torque component ($q$-axis) of current in a drive, thereby leading to wobbles and fluctuations in motor speed and phase currents. The voltage sensor fault can negatively affect the performance of the flux observer and the estimation of motor parameters. In summary, all of these consequences resulting from the erroneous feedback due to sensor failures could lead to degraded control performance or even drive system shutdown.

Traditional sensor fault monitoring methods can be divided into model-based and signal-based methods \cite{gao2015survey}. Model-based methods aim to evaluate and monitor the difference between the measured output of the actual system and the output generated by the model, which is typically obtained using state observers or MRAS-based approaches for continuous estimation of rotor speed and phase currents. While the model-based methods are fast and independent of operating conditions, their performance is also highly dependent on the accuracy of the model and its parameters. For signal-based methods, the diagnostic process is based on the real-time evaluation of fault signatures obtained by manual feature extraction. For example, the average normalized current value, the magnitude of the harmonic frequency components, and the asymmetry between the phase currents can all be used to identify current sensor faults. Despite their independence from system parameters and models, these signal-based methods are highly sensitive to the load conditions of the drive system and often require some expert domain knowledge to manually select the useful features.

Based on recent advancements, ML-based methods also have the potential to become a promising alternative to the condition monitoring of sensors for motor drive applications. A summary of currently available publications on this topic using ML techniques is presented in TABLE \ref{tab:drive_sensor_monitoring}. While only a few publications exist and most of them are also fairly recent compared to model-based and signal-based approaches, it is reported that ML-based solutions demonstrate some superior condition monitoring performance with higher generalization capability and increased robustness \cite{gou2018intelligent, gou2020online, xia2021learning, jankowska2022design}. However, one potential concern still remains related to the model size/parameters and the corresponding inference time. This range can vary from \SI{10}{ms} to a few seconds and some of which may not be fast enough to meet the requirement of certain drive applications.

\section{Reinforcement Learning-Enabled Next Generation Electric Machine Drives}\label{sec:Future_drives}
Despite the widespread applications of RL in AlphaGo, robots, and self-driving cars, RL has only recently been introduced to the control of electric machine drives \cite{schenke2019controller, traue2020toward, balakrishna2021gym, hanke2020data1, hanke2020data2, schenke2021deep, book2021transferring, schindler2019comparison, bhattacharjee2020advanced, el2020adaptive, alharkan2021optimal}. Similar to the vision of self-driving cars where a car can drive itself and take its passengers to their desired destinations, RL-enabled electric machine drives are expected to meet various performance requirements and efficiency specifications by automatically learning their optimal control policies via direct interactions with the actual motors. This entire workflow can be completely automated and does only require minimum human design effort as well as  a priori model knowledge. 
%

%
%
\begin{figure}[!t]
    \centering
    \subfloat[]{\includegraphics[width=3.2in]{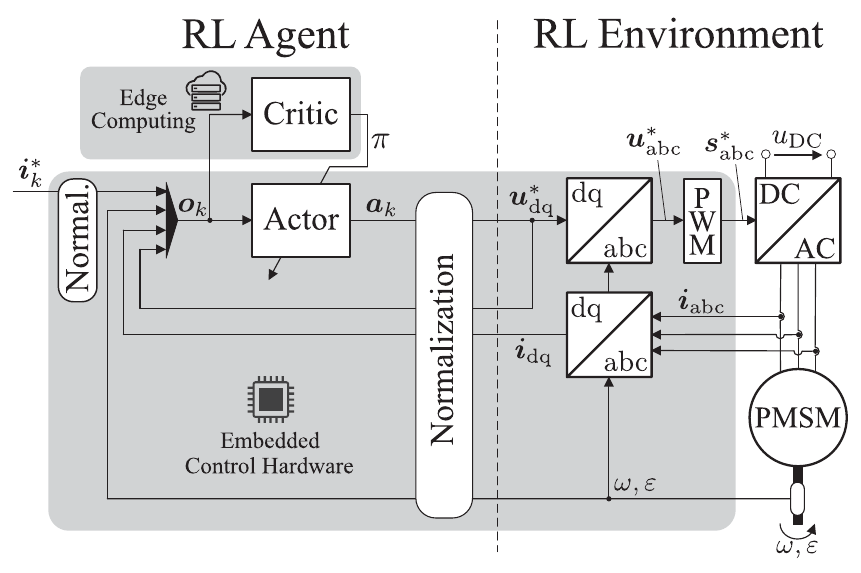}} \\
    \subfloat[]{\includegraphics[width=3.2in]{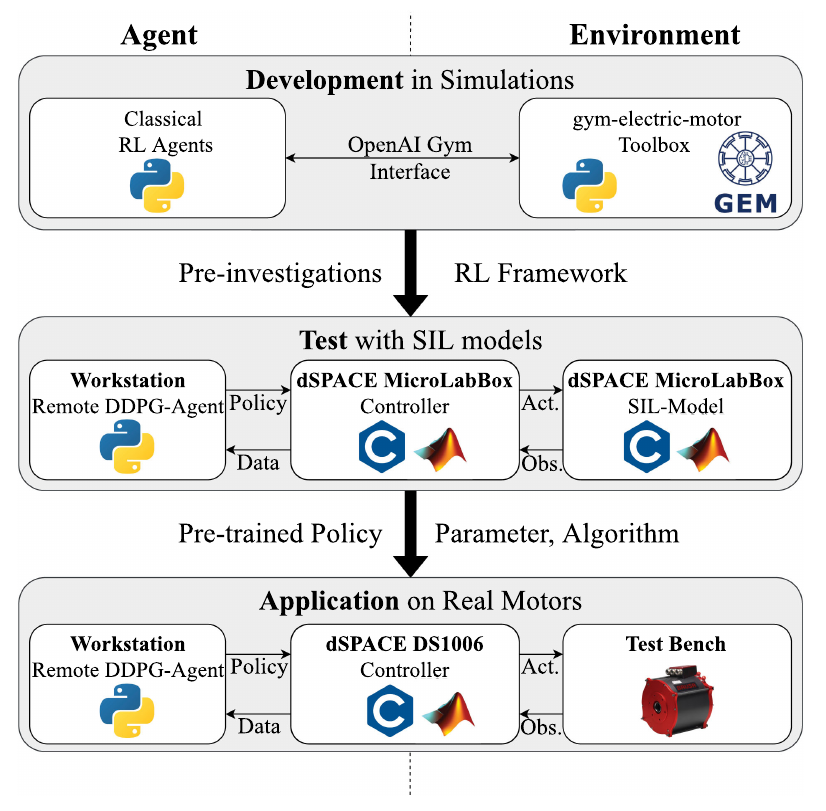}} 
    \caption{(a) Simplified schematic of the overall reinforcement-learning-based control and drive system structure partitioning the agent and environment, and (b) setup of the development process including the online RL remote rapid control prototyping toolchain \cite{book2021transferring}.}
    \label{fig:RL_motor_drive}
\end{figure}

To make this RL-enabled motor control scheme a competitive alternative to classical methods, many exemplary works have been presented over the course of the past few years exploring the boundary and tackling the unsolved problems \cite{schenke2019controller, traue2020toward, balakrishna2021gym, hanke2020data1, hanke2020data2, schenke2021deep, book2021transferring}. The first proof-of-concept of the RL-based current control in a PMSM drive is presented in \cite{schenke2019controller}, which successfully validates the basic design architecture shown in Fig. \ref{fig:RL_motor_drive}(a) and underlines the potential of the data-driven controller design. To accelerate the development and training of RL agents for electric motor control, an open-source gym-electric-motor Python toolbox is published in \cite{traue2020toward, balakrishna2021gym}, which contains models of different DC and three-phase motor variants for easily accessible simulation. This package can be readily used to compare the trained RL agents with other state-of-the-art control approaches. For the same purpose, a data set consisting of about 40 million data points is recorded at a test bench for a 57-kW PM machine drive and is published on Kaggle \cite{hanke2020data1, hanke2020data2}. A deep Q-learning (DQN) direct torque controller is further implemented for PM machines by aligning the limited number of distinct switching states of voltage source inverters and DQN's finite control set framework.

More recently, another important step is accomplished towards introducing RL to the control of physical motor drives, which involves the complete workflow of transferring an RL controller from offline simulation to online training and inference on a real motor drive system, as illustrated in Fig. \ref{fig:RL_motor_drive}(b) \cite{book2021transferring}. In order to outsource computational heavy RL computations, edge computing based on an internet of the things framework is utilized. Consequently, only the control policy inference must be calculated in real time on the embedded controller while the actual RL training algorithms are calculated in an asynchronous fashion using dedicated computing resources. It is further envisioned that such an implementation will also be possible for low-cost applications in the future using typical system-on-chip (SoC) embedded hardware with FPGA, as will be detailed in the next section of implementing ML-based motor drives in embedded systems. Furthermore, \cite{Schenke2023} extends the scheme with an online safeguarding method to prevent unsafe drive operations due to random exploration actions. By doing so, online torque control learning can be accomplished in less than 10 minutes on a real-world system. 

Besides the pioneering work mentioned above, readers are also referred to other state-of-the-art literature employing RL to PM machine \cite{schindler2019comparison, bhattacharjee2020advanced, el2020adaptive} and switched reluctance machine drives \cite{alharkan2021optimal} for more details.

\section{Implementing Machine Learning-Based Electric Machine Drives in Embedded Systems}\label{sec:Embedded}
\subsection{A Brief History of Embedded Systems for Electric Machine Drives}
Due to the lack of a suitable ANN application-specific integrated circuit (ASIC) or FPGA in the 1990s, the experimental validation of the first ML-based control algorithms for electric drives has been performed on microcomputers or microprocessors, focusing on available parallel architectures. For example, the first ANN-based current controller to identify and control the induction machine dynamics presented in \cite{wishart1995identification} uses an \SI{25}{\MHz} INMOS T800 transputer with a 32-bit integer processor that runs in parallel with a 64-bit floating-point unit on a single chip \cite{burton1998implementation}. Due to hardware limitations, the final attainable sampling rate is \SI{500}{\Hz} with a two-layer ANN of 8 inputs, 12 hidden nodes, and two outputs. It is reported that the stator currents will show signs of growing instability with the increase of its electrical frequency until reaching a point as low as \SI{1.27}{\Hz}, where the ANN controller behaves wildly. Therefore, it is suggested that the \SI{500}{\Hz} prototype ANN current controller must be increased by an order of magnitude, and higher speeds of computation will be required from the hardware.

Additionally, two model reference adaptive speed neural controllers proposed in \cite{shyu1998model, chen2002model} are implemented in x86 microcomputers with only a \SI{500}{\Hz} sampling rate,  
though they are still shown to compare favorably against the benchmark PI controllers during transients \cite{chen2002model}. Furthermore, \cite{mohamadian1997dsp} and \cite{mohamadian2003novel} provide an exemplar study that runs an ANN-based current controller and the rest of the indirect FOC control on a Texas Instruments TMS320C30 DSP. Despite implementing certain optimization strategies such as performing the hyper tangent sigmoid function by a look-up table, the final attainable sampling frequency is still only \SI{1}{\kHz} due to hardware limitations, though researchers always tend to use the best available parallel hardware.

During this time, a variety of algorithmic approaches are also proposed in \cite{burton1996linear, burton1997identification, burton1998reducing, burton1999high} to accelerate the continual online training and to enable a sampling frequency of at least \SI{10}{\kHz} for modern electric machine drives. These acceleration methods include efficient parallelization methods such as output separation and tandem parallelization \cite{burton1996linear}; the random weight change algorithm to replace the conventional backpropagation for online training \cite{burton1997identification, burton1999high}; as well as various techniques to reduce the computational demand \cite{burton1998reducing}.
%

%
With the evolution of hardware capabilities in the new century, ML-based controllers for machine drives have been advanced to execute at or above the desired switching frequency. As presented in \cite{restrepo2004practical, restrepo2004ann, restrepo2007induction}, all of the computations related to the same two-layer ANN proposed in \cite{burton1998reducing} are now able to run at \SI{10}{\kHz} to identify the system dynamics within \SI{1}{\ms} using the pre-trained weights. The ML-based controller is deployed on a \SI{333}{\MHz} Analog Devices ADSP-21369 DSP that is capable of executing at 2 giga floating-point instructions per second (GFLOPS). An interface card is also used to host two FPGAs in charge of handling the high-speed parallel data coming from the data acquisition system \cite{restrepo2007induction}. In \cite{lin2010fpga}, a field-oriented control PM linear machine drive is implemented on a \SI{24}{\MHz} Xilinx XC2V1000 FPGA with a switching frequency of \SI{15}{\kHz}. In addition, ML models have also been on FPGAs integrated with the  National Instruments CompactRIO controller for a two-mass electric drive system. Specifically, a multilayer perception network is implemented for the speed estimation \cite{orlowska2011fpga} and an ADALINE model is implemented as a speed controller \cite{kaminski2012fpga}.

Similarly, while the validation of ML-based flux observers is only carried out in simulation in the 1990s \cite{toh1994flux, theocharis1994neural, simoes1995neural, ba1997field, marino1999linear, pinto2000neural}, the evolution of hardware platforms, especially FPGAs, has further advanced the implementation and validation of ML-based observers on the hardware. For example, a flux observer with two cascaded ANNs has been implemented in \cite{zhang2008stochastic} using a single XC3S400 FPGA from Xilinx, and validation of the proposed FPGA controller is performed on a hardware-in-the-loop (HIL) test platform using a real-time digital simulator with a \SI{50}{\micro\second} time step. Efficient inference of ML modes is also achieved in \cite{soares2006field} when deploying a stator flux-oriented induction machine drive on a Stratix 2 FPGA by Altera. Specifically, the computational time for synthesizing the SVPWM using a three-layer ANN and for estimating the stator flux using a three-layer RNN are only \SI{1.7}{\micro\second} and \SI{1.0}{\micro\second}, respectively.

However, it should be noted that most of the aforementioned ML models still have shallow structures with only one or two hidden layers of a neural network with dozens of neurons at most, and the inference of significantly larger neural networks can be achieved using today's mainstream FPGAs. A recent study that seeks to investigate the boundary conditions to use multilayer perception (MLP) networks in motor control applications has been performed on the Xilinx Zynq UltraScale ZU9EG FPGA with 2,520 DSP slices \cite{schindler2020real}, which reveals that the inference of an MLP with 3 hidden layers and 64 neurons in each layer can be completed in \SI{7.36}{\micro\second} with 32 parallel neuron control units. It is also reported from the proposed implementation that it is more efficient for the inference of deeper MLPs (more hidden layers) compared to MLPs with a high number of neurons per layer and fewer hidden layers. An overview of recent achievements in the area of FPGA and GPU-based implementations for reinforcement learning is provided in \cite{rothmann2022survey}.

Besides widely-used embedded systems based on DSPs and FPGAs, some rapid control prototyping platforms are also leveraged to deploy ML-based motor control applications. For example, \cite{kaminski2020nature} proposes an ML-based induction machine drive composed of a parallel combination of the classical PI structure and the radial basis function neural network on a dSPACE DS1103 card. In addition, the inference of ANN and the rest of the vector control algorithm are also implemented on a dSPACE DS1103 in \cite{fu2015novel} with a sampling frequency up to \SI{10}{\kHz}. The hardware experiments further reveal that when compared with the PI controllers, the ANN-based controllers can achieve much better current tracking performance with a low PWM switching frequency of \SI{4}{\kHz}, which further yields possibilities to improve the motor drive efficiency by lowering its switching loss.


\subsection{Selecting Appropriate Embedded Systems for ML-Based Electric Machine Drives}
Although various ML-based electric machine drives have been successfully implemented in embedded systems with DSPs \cite{vas1998dsp, rubaai2008dsp, demirtas2009dsp, mohamadian1997dsp, suetake2010embedded}, FPGAs\cite{soares2006field, le2009neural, lin2010fpga, quang2014fpga, kaminski2012fpga, orlowska2011fpga, monmasson2011fpgas}, and embedded GPUs (\cite{mohamed2019neural, truong2022light, lehment2022interplay, gawde2022multi, blaha2022real}) during the past 30 years, most of them have rather shallow network structures and slow PWM cycles. Fortunately, the development of hardware platforms for parallel computing, including GPUs, FPGAs, and TPUs, has significantly promoted the fast evolution and deployment of ML algorithms for industrial applications in recent years. A clear example is the currently very active domain of perception algorithms for advanced driver-assistance systems and autonomous driving. Based on the parallel properties inherent in such deep neural networks applied to electric machine drives, an FPGA-based or GPU-based implementation also seems promising and is highly recommended in \cite{monmasson2021system}.

Due to their intrinsic architecture, however, GPUs are only efficient for processing data with large batch sizes that fit into the scope of CNNs. On the other hand, the control of electric machines will always utilize a handful of real-time measured signals that have vastly different data representations than raw pixel data processed primarily by CNNs. As such, GPUs may not be the most appropriate platforms for electric machine drives that require ultra-low latency and high interfacing flexibility -- though both of which are the strengths of FPGAs. Therefore, we'll focus on FPGAs for the remainder of this section with detailed discussions provided as follows.

For ML-based high-performance electric machine drives, an ultra-low latency in the order of microseconds will be needed because the control frequency of which is typically in the range of \SI{10}{\kHz} to \SI{40}{\kHz}, hence the maximum available calculation time for each control loop is $t_\mathrm{c} =$ \SI{25}{\micro\second} to $t_\mathrm{c} =$ \SI{100}{\micro\second}. Excluding the time needed for ADC sampling, signal scaling/filtering, software-based protection logic, etc., the available time for the inference of deep neural networks has to be always lower than a full control cycle. Furthermore, machine drives will also need to interface with many different categories of sensors to properly perform the control, estimation, and monitoring of electric machines for different industrial applications.
%



In addition to the low latency and high interfacing flexibility discussed earlier, there are also many other advantages of using an FPGA for the inference of deep neural networks, as presented in TABLE \ref{tab:platform_comparison}. We'll also elaborate on how these advantages are particularly relevant to motor control applications as follows:


%
%
\begin{table}[!t]
    \caption{Performance comparison between CPU, GPU, and FPGA for the inference of neural networks \cite{tao2020challenges}.}
    \centering
\begin{tabular}{lccc}
\toprule
\rowcolor[HTML]{FFFFFF} 
                        & CPU & GPU                   & FPGA                      \\ 
\midrule
\rowcolor[HTML]{EFEFEF} 
Throughput              & Lowest     & Highest          & High                     \\
\rowcolor[HTML]{FFFFFF} 
Latency                 & Highest    & Medium           & Lowest                   \\
\rowcolor[HTML]{EFEFEF} 
Power                   & Medium     & Highest          & Lowest                   \\
\rowcolor[HTML]{FFFFFF} 
Energy Efficiency       & Worst      & Medium           & Best                     \\
\rowcolor[HTML]{EFEFEF} 
Device Size             & Small      & Large            & Small                    \\
\rowcolor[HTML]{FFFFFF} 
Development             & Easiest    & Easy             & Hard                     \\
\rowcolor[HTML]{EFEFEF} 
Library Support         & Sufficient & Sufficient       & Limited                  \\ 
\rowcolor[HTML]{FFFFFF} 
Flexibility             & Limited    & Limited          & Flexible                  \\
\bottomrule
\end{tabular}
\label{tab:platform_comparison}
\end{table}
\begin{figure*}[!t]
\centering
\includegraphics[width=6.4in]{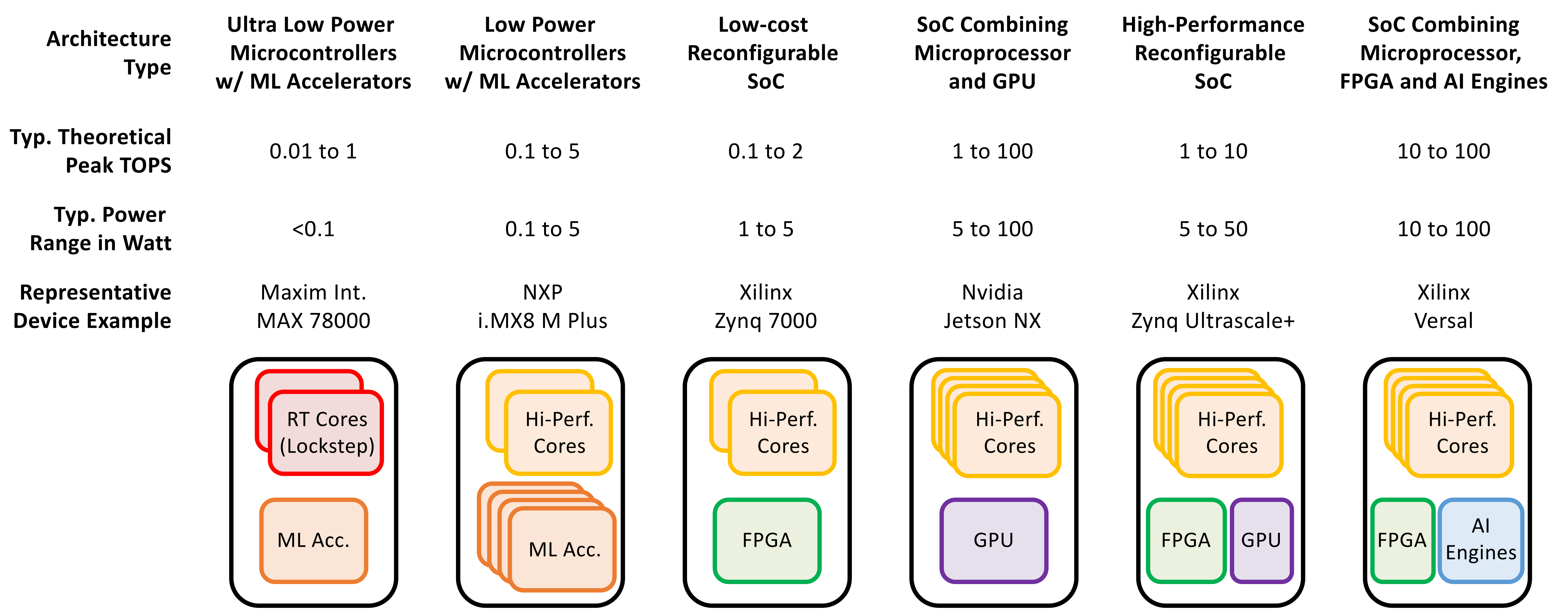}
\caption{Overview of relevant embedded platform types in the market, illustrating a simplified block diagram of their topology, providing indicative ranges for the typical theoretical peak TOPS performance and power consumption for each type and one representative device example, figure adapted from \cite{dendaluce2019efficient}.}
\label{fig:embedded}
\end{figure*}
\begin{enumerate}
    \item[1)] \emph{Low latency}: Latency is important in the inference of neural networks as it is directly tied to their real-time performance. FPGAs offer clear advantages over GPUs and CPUs with lower latencies, which is a prerequisite for applications that run inference in real-time, such as the control of electric machines (including online RL training). 
    This advantage can be attributed to the fact that FPGAs can be configured to directly access peripheral hardware components, such as sensors or input data sources. Directly combining this with implementations for the required preprocessing in the FPGA fabric provides a very high bandwidth and much lower latency. On the other hand, the communication between GPUs and hardware components is less efficient, since a standard bus (USB or PCIe) is typically required to access the hardware, and a host system (or an embedded CPU) needs to be employed \cite{tao2020challenges}. Furthermore, based on their architecture, requiring a high number of threads running in parallel, GPUs can provide high bandwidth only at the cost of high latency since they are only efficient for large batch sizes.
    As a qualitative example, \cite{schindler2020real} shows that the latency of a reinforcement learning-based motor control application can be reduced to as low as \SI{7.36}{\micro\second} on FPGAs, which is sufficient for a control frequency of 100 kHz. Specifically, the deployed neural network has 9,224 variables and the inference is performed using 32 DSP-slices, which are offered by the programmable logic part of the Xilinx FPGA to efficiently implement multiplications and multiply-accumulate operations. Although the number of DSP-slices is a limited resource on FPGAs, it seems there's still big headroom for FPGAs to run inference on deeper and larger neural networks for motor controls. For example, the current implementation in \cite{schindler2020real} uses 32 DSP-slices to get to a point where the latency is below \SI{10}{\micro\second}, while the low-end Xilinx Zynq-7020 offers 220 DSP-slices \cite{Zynq_7000} and Xilinx UltraScale ZU2EG offers 240 DSP-slices \cite{Zynq_UltraScale+}. 
    \item[2)] \emph{Excellent interfacing flexibility}: FPGAs can be reprogrammed for different functionalities and data types \cite{fpga_vs_gpu}. They also excel at handling data input from multiple sensors, such as current sensors, voltage sensors, thermocouples, encoders/resolvers, and accelerometers. These features make FPGAs very flexible when optimizing hardware acceleration of ML inference for electric machine drives.
    \item[3)] \emph{High throughput}: Based on the tightly-coupled SoC architecture, FPGAs can deliver high throughput by optimizing hardware acceleration of ML inference in the programmable logic (PL) part and other non-critical functions in the processing system (PS). Additionally, they also have the capability of hardware software co-design to achieve optimized balancing between the two. These desirable features could bring matched throughput with end-to-end applications that are able to deliver significantly better performance than fixed-architecture AI accelerators such as GPUs. That's because with a GPU, the other non-critical functions of the application must still run in software without the performance or efficiency of custom hardware acceleration. 
    %
    \item[4)] \emph{Affordable cost}: Large GPUs can be excessively costly to be considered suitable for many electric drive applications, including home appliances, pumps, fans, or even electric vehicles, while FPGAs are often more affordable. 
    By integrating additional capabilities onto the same chip thanks to its SoC architecture, designers can also save on cost and board space. In addition, FPGAs have long product life cycles, measured in years or decades. This characteristic makes them ideal for use in industrial, defense, medical, and automotive markets as it further reduces maintenance costs. Despite the costs of FPGAs being still expensive when compared with the standard micro-controllers that host classical FOC/DTC motor control algorithms, the reconfigurable SoC can offer an integrated and much simpler design of the software program and the hardware FPGA image. More importantly, there's a great potential for using ML-based methods in terms of quick exploration and domain adaptation on motor control over existing methods than run on these ultra-low-cost micro-controllers.
    \item[5)] \emph{Low power consumption}: With FPGAs, designers can fine-tune the hardware according to the application to help meet energy efficiency requirements. FPGAs can also provide a variety of functions to improve the energy efficiency of the chip. It’s possible to use a portion of an FPGA for a function instead of the entire chip, allowing the FPGA to host multiple functions in parallel and the ability of dynamic reconfiguration \cite{fpga_vs_gpu}.
\end{enumerate}
\begin{figure}[!t]
\centering
\includegraphics[width=3.6in]{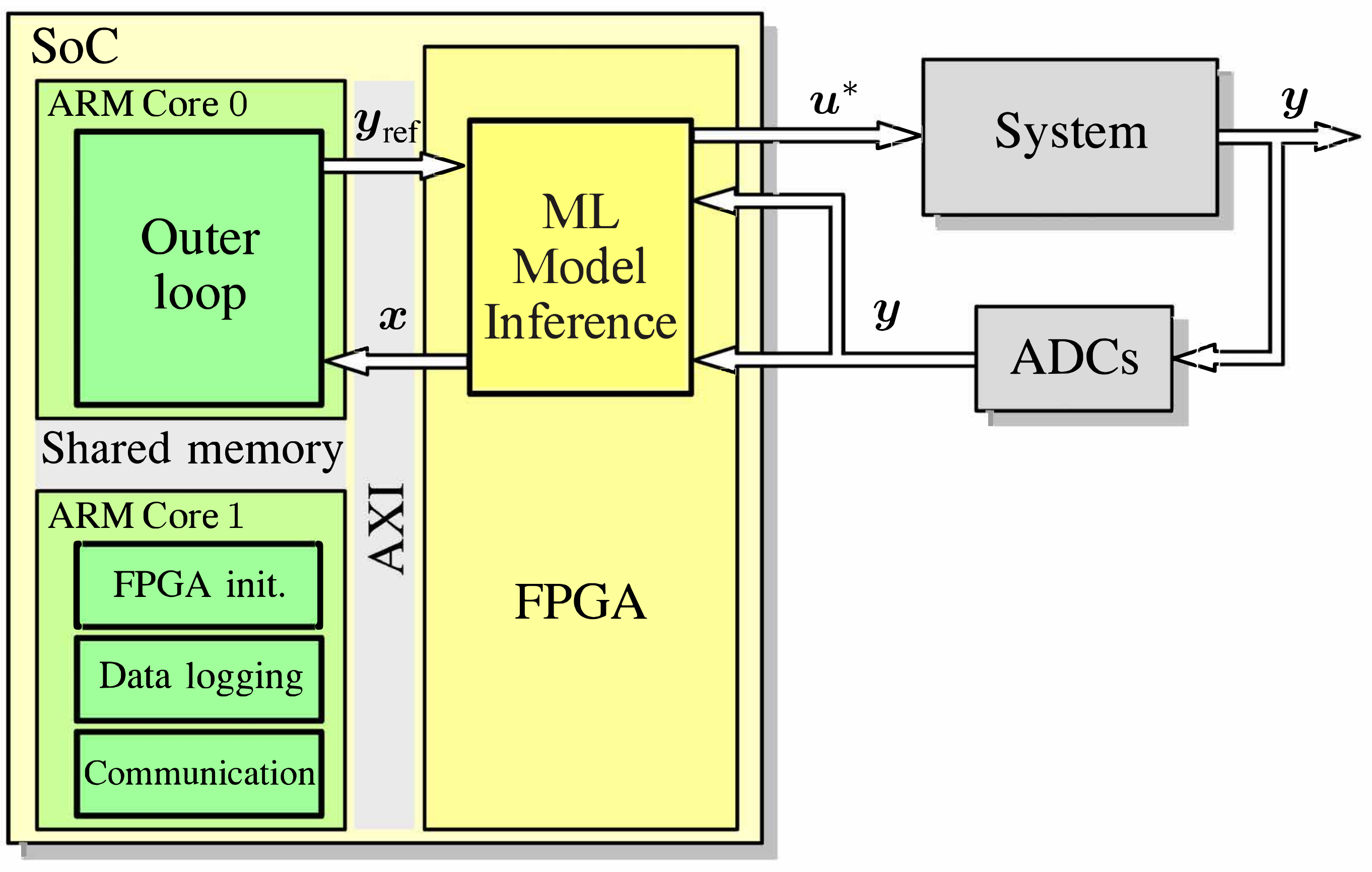}
\caption{FPGA-based SoC structure for the inference of ML models for motor control applications, figure adapted from \cite{karamanakos2020model}}.
\label{fig:SoC_image}
\end{figure}

Based on the aforementioned comparisons, it can be concluded that FPGAs, especially those based on the SoC architecture, are among the most promising digital technologies for implementing ML-based smart controllers in electric drives. Specifically, the reconfigurable SoC consists of memory, microprocessors, analog interfaces, an on-chip network, and a programmable logic block. Additionally, heterogeneous multi-processing SoC (MPSoC) architectures offer better performance in terms of power and performance when compared with monolithic cores \cite{bahri2013hardware}. Examples of such a new class of reconfigurable SoCs are the Xilinx All-Programmable Zynq, the Altera reconfigurable SoC, and the Actel/Microsemi M1 \cite{trimberger2015three}. Fig. \ref{fig:embedded} gives an overview of the different architectures that are available on the market, providing an indication of typical performance and power ranges. In 2018, Xilinx also launched a new programmable chip architecture called the adaptive compute acceleration platform (ACAP), a re-programmable multi-core compute architecture with new dedicated AI engines integrated into the device. With this heterogeneous approach, the architecture goes beyond the capabilities of current reconfigurable SoCs and can even be modified dynamically in milliseconds during operation to meet changing workload requirements \cite{versal_news}. The latest Xilinx edge Versal VE1752 is now shipping out to customers \cite{versal} and it could become a favorable embedded platform for next-generation motor drive applications.

\subsection{Implementing Machine Learning-Based Motor Control in FPGAs}
Fig. \ref{fig:SoC_image} depicts a simplified example of the implementation of an ML-based motor control  algorithm on a dual-core reconfigurable SoC. First, the measurements are read from the ADCs and processed by digital filters implemented in the FPGA. Subsequently, the inference of neural networks is executed in the FPGA that also estimates the current state $x(k)$. The reference command (torque, speed, or position) $y_{\mathrm{ref}}(k)$ is provided by an outer control loop that runs on the ARM Core 0. The interface between Core 0 and FPGA is realized by the integrated advanced extensible interface (AXI). The other depicted ARM Core 1 is generally not part of the control loop, but it is responsible for many ``housekeeping'' tasks, such as data logging, communication with other systems and users, and the initialization of the FPGA, which includes all the libraries, all the tenants, the real-time operating system (RTOS), drivers, and application programming interfaces (API), etc.


%
%
%

However, it is also worthwhile to mention that FPGAs can be difficult to program as they require significant hardware design expertise or long learning curves for optimal use, and the task of converting sequential, high-level software descriptions into fully optimized, parallel hardware architectures is tremendously complex \cite{gschwend2020zynqnet}. This limitation is only becoming more profound when deploying ML algorithms with a deep structure and a large number of parameters. 
Fortunately, instead of starting from scratch, there are many different tools and customized environments to streamline this process. To provide some examples, we'll present some potential ways to deploy a trained ML-based controller for electric drives in the FPGA.

\subsubsection{PYNQ -- Python Productivity for Zynq}

PYNQ is an open-source project from Xilinx that makes it easier to use Xilinx platforms by using the Python language and libraries \cite{Pynq}. Compatible with Zynq, Zynq UltraScale+, Zynq RFSoC, and Alveo accelerator boards, the PYNQ platform improves the productivity of designers already working with Zynq, and it reduces the barrier to entry for users with limited experience in hardware design. Fig. \ref{fig:Pynq} illustrates the general concept of the PYNQ framework consisting of three layers:
\begin{enumerate}
    \item[$\bigcdot$] \emph{Upper Layer (Applications)}: The upper layer of the PYNQ stack enables user interaction using one or more Jupyter Notebooks, which are hosted on Zynq’s Arm processors, also known as the processing system. Custom functionalities specific to each application can be created by writing Python code and using many open-source Python libraries. In addition to developing software-based functionality running on the PS, Python code within the notebook can also offload processing to hardware modules operating on the PL \cite{pynq_book}. Interaction with hardware is achieved using the Python APIs and drivers that are provided as part of the PYNQ framework. The programmer’s experience of using hardware blocks is therefore very similar to calling functions from a software library — a software developer can call a hardware block without any need to understand the internals of the hardware design. 
    \item[$\bigcdot$] \emph{Middle Layers (Software)}: In the middle layer, the PYNQ framework includes a Linux-based OS, bootloaders to initiate system start-up, a web server to host Jupyter notebooks, and a set of drivers for interacting with elements of the Zynq hardware system. Thus, the design effort of developing common software elements of an embedded system is significantly reduced, and new users are expected to get started quickly with Zynq. 
    \begin{figure}[!t]
    \centering
    \includegraphics[width=3.4in]{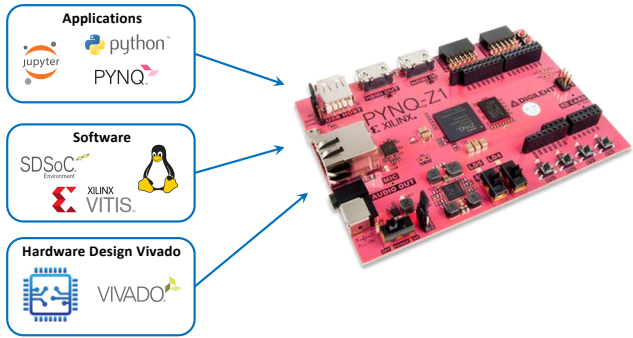}
    \caption{PYNQ -- an open-source project from Xilinx that features an easy software interface and framework for rapid prototyping and development \cite{Pynq}.}
    \label{fig:Pynq}
    \end{figure}
    %
    %
    \item[$\bigcdot$] \emph{Lower Layer (Hardware)}: The bottom layer of the stack represents a hardware system design, which would normally be created in Vivado requiring significant hardware design expertise. In PYNQ, however, hardware system designs are often referred to as overlays and they can be used in a manner analogous to software libraries. Specifically, PYNQ provided a base hardware system with an aspect of generality that includes almost all modules in the PYNQ board for flexible reuse, such as interfacing blocks for DMA, audio, video, I2C, and components from logic tools.
    Neural network accelerators can then be implemented through such overlays, as presented in \cite{hao2017implementation}, which deployed a recurrent neural network language model for speech recognition.
\end{enumerate}
%
%
%

However, it should be noted that new accelerators have to be developed from scratch within the PYNQ framework, and similar to FPGAs in general, it is mostly limited to the inference of a neural network, and the online learning through back-propagation is usually difficult to implement on such low-cost FPGAs in general due to limited resources. Alternatively, some on-device learning approaches that do not rely on back-propagation for training have been proposed for FPGAs \cite{tsukada2020neural, ito2021device, watanabe2021fpga}.

\subsubsection{Matlab HDL Coder and Xilinx System Generator (XSG)}
HDL Coder provides a workflow advisor that automates the programming of Xilinx, Microsemi, and Intel FPGAs \cite{hdl_coder}. Specifically, it can generate portable, synthesizable Verilog and VHDL code from over 300 HDL-ready Simulink blocks, MATLAB functions, and Stateflow charts. With HDL Coder, programming FPGAs for ML-based motor control applications can be achieved at a high level of abstraction, and the generated HDL code can be imported and compiled into customized IP cores using the Intel Quartus or the Xilinx Vivado Design Suite.

Besides the HDL Coder, Xilinx also developed its own Xilinx System Generator (XSG) that adds Xilinx-specific blocks to Simulink for system-level simulation and hardware deployment. We can also integrate System Generator blocks with native Simulink blocks for HDL code generation on the desired neural network structure. In \cite{schindler2020real}, for example, 
the VHDL code for two multi-layer perceptions (MLP) neural networks is also generated by the HDL Coder. 

By adopting such a model-based workflow utilizing the HDL Coder, the proper functioning of the system can be first examined by simulation and co-simulation in Matlab, then the block design is integrated into the FPGA architecture in the form of an IP core. This workflow is very convenient for high-level integration of various IP blocks created using the Matlab/Simulink graphical interface, especially for those who are not familiar with hardware description languages such as VHDL and Verilog. Also, the debugging and verification of HDL designs become easy and flexible with the Simulink toolbox, though the performance and resource utilization of such toolboxes may not yield the optimal design compared to experienced FPGA designers.


\subsubsection{Deep Learning Processor Unit (DPU)}
Besides the high-level synthesis (HLS) tool that can compile deep learning C/C++ code for programmable logic in the hardware \cite{hassan2020implementation}, Xilinx also developed The Deep Learning Processor Unit (DPU) intellectual property (IP) core that can be integrated into the programmable logic of selected Zynq-7000 SoC, Zynq UltraScale+ MPSoC, and Versal AI edge devices with direct connections to the processing system. Specifically, this DPU is a programmable engine dedicated to convolutional neural networks. This unit includes the register configuration module, the data controller module, and the convolution computing module. The DPU has a specialized instruction set, which allows the DPU to work efficiently on many convolutional neural networks, including VGG, ResNet, GoogLeNet, YOLO, SSD, MobileNet, FPN, etc. The figure below shows an example system block diagram with the Xilinx UltraScale+ MPSoC using a camera input. The DPU is integrated into the system through an AXI interconnect to perform deep learning inference tasks such as image classification, object detection, and semantic segmentation \cite{dpu}.
\begin{figure}[!t]
\centering
\includegraphics[width=3.4in]{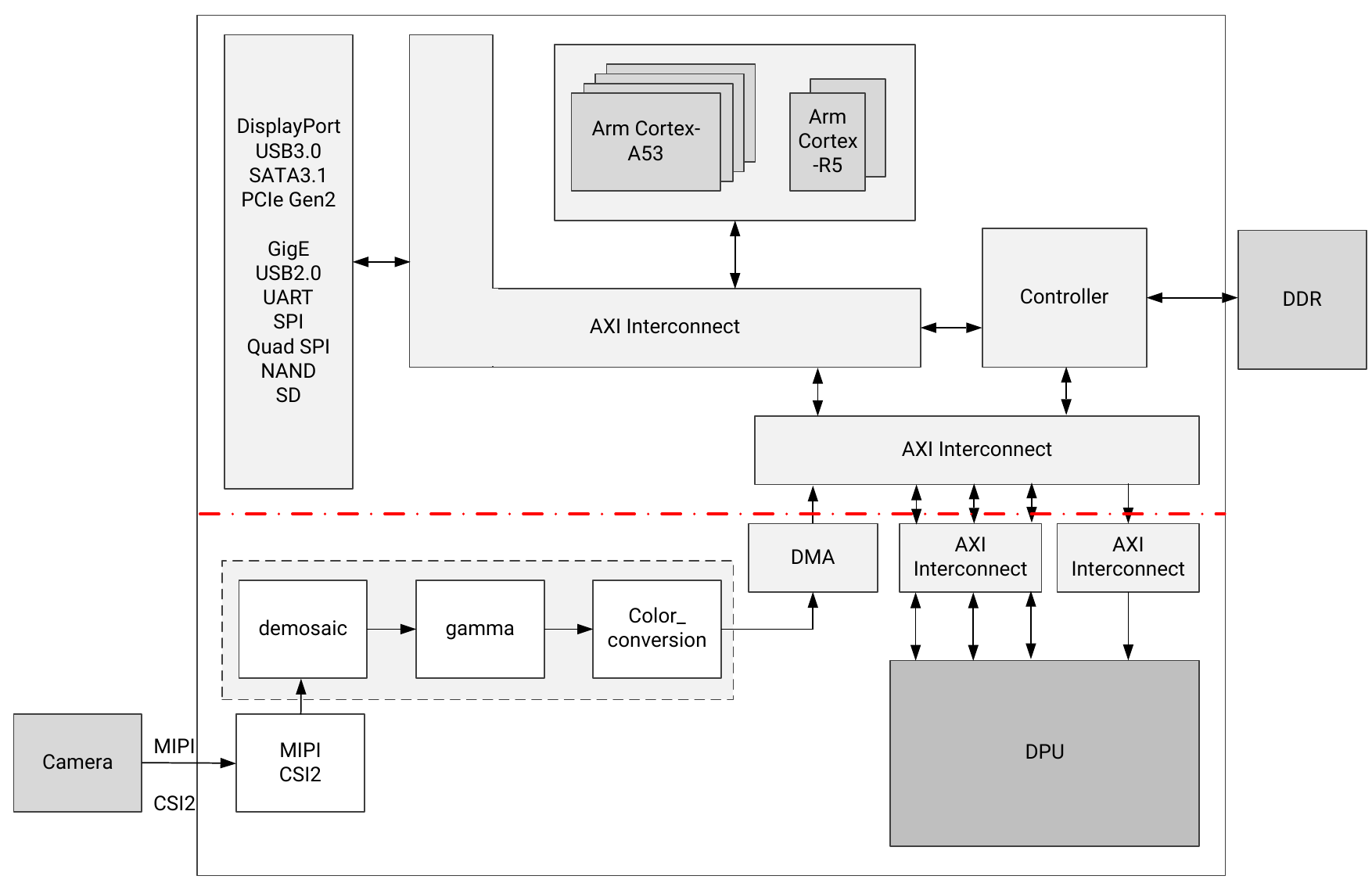}
\caption{Example system with an integrated deep learning processor unit (DPU) \cite{dpu}.}
\label{fig:dpu}
\end{figure}

This Xilinx DPU IP module is provided at no additional cost with the Xilinx Vivado Design Suite. However, it should be noted that as a CNN IP core, DPU is highly tailored for computer vision and image recognition-related applications, where users are expected to prepare the instructions and input image data in the specific memory address that DPU can access. Although CNNs are seldom used to tackle control tasks of high complexities - such as electric machine drives, the convolutional layers can often be deployed as a part of the reinforcement learning algorithms. For example, in order to learn good policies with just pixel inputs, the authors of the deep deterministic policy gradient (DDPG) algorithm used 3 convolutional layers to provide an easily separable representation of state space \cite{lillicrap2016continuous}. 

However, one should also note that \cite{lillicrap2016continuous} learns from raw pixels and, therefore, processes image data where CNNs shine. As discussed earlier, electric machine drives utilize a completely different data representation than what CNNs are primarily used to process. Since CNNs are really costly in computation, the prospect of their applications tailored for electric drives on low-cost embedded platforms such as FPGAs for low-dimensional control tasks is somewhat unclear. Nevertheless, if CNNs are selected to accomplish certain motor control tasks, we can still benefit from this DPU IP core by taking advantage of its built-in convolutional layers and integrating them with other layers of the neural network designed in custom IP cores.

\subsection{Others}
The actual hardware design of FPGAs can be performed by combining any of the methods mentioned above. In addition. some advanced high-level synthesis (HLS) tools, such as the Auto-HLS \cite{hao2019fpga}, can be used to directly generate synthesizable C code of the ML models and to conduct latency/resource estimation and FPGA accelerator generation. 

The framework of FINN \cite{umuroglu2017finn} can also be adapted to build fast and flexible FPGA accelerators by reducing the weights and activations of ML models for motor drive applications to low bit width or even binary values. This method is especially well-suited for CNNs that contain significant redundancy, and a similar motor drive performance is expected against the original ML model without adapting to the FINN framework.

In addition to embedded control systems, commercial rapid control prototyping (RCP) systems have also been used in deploying ML-based motor control algorithms. Such systems include the dSPACE MicroLabBox and DS1006MC \cite{book2021transferring}, which implement a deep deterministic policy gradient algorithm that learns the current control policy for a PM motor. Moreover, open-source software and hardware RCP systems, such as UltraZohm \cite{9590016} or AMDC \cite{severson-group}, can contribute to distributing open ML-based drive control and monitoring solution.

\section{Future Challenges and Trends}\label{sec:Conclusions}

This paper provides a comprehensive state-of-the-art review of ML-based solutions addressing the control and monitoring of electric drives. Despite the continued progress of relevant publications in this field, there are still some unresolved issues that need to be addressed as future work:
\begin{enumerate}
    \item[$\bigcdot$] \emph{Development effort}: Although FPGAs can offer better energy efficiency, connectivity, and flexibility, one major challenge of using FPGAs is the engineering effort in development. Unlike GPU development which requires only software engineering skills, the development of FPGAs requires hardware configuration skills as well. 
    The complexity of implementing ML models on FPGAs makes their manual design processes very time-consuming, even for a seasoned FPGA engineer. In addition, although many researchers have focused on ML inference on FPGAs, very few research papers have explored their training on FPGAs, or how to optimize the architecture design on FPGAs for training. This is particularly needed for deploying RL algorithms to motor control applications as the core of RL is to be able to have the agent interact with the environment in real time and learn a policy (control law) in a trial-an-error way. Therefore, an automated design workflow from the RL's neural network architecture to the hardware design is necessary to enable efficient and effective training of RL control on FPGAs (i.e., to not only utilize the FPGA for policy inference but also for online policy learning). 
    If an effective automated design workflow is developed, researchers and engineers can quickly develop various ML models for motor control applications without the need to possess deep knowledge about hardware design.
    \item[$\bigcdot$] \emph{Application effort}: ML is data-hungry and normally one needs to train an ML model individually for each drive system at an expensive test bench, hence the speedy transfer of an ML method between different applications is an issue for the industrial mass production usage. This issue can be addressed from both the software and the hardware emulation perspectives. In terms of software, some ML algorithms are specifically designed to enable transfer learning with strong domain adaptation capabilities. Additionally, the hardware platform of different electric drive systems can be emulated in the hardware-in-the-loop environment, making it a lot easier to collect a sufficient amount of emulated data to train ML models for any industry application.
    \item[$\bigcdot$] \emph{Safety}: Since ML is always subject to some kind of stochastic learning, a method's output should be also considered a stochastic one. Therefore, the probability of failing is intrinsic to an ML model, which can cause trouble if an ML technique produces outliers for estimation or control action. As a result, there could be negative impacts on the behaviors of mechatronic systems, thereby compromising their prospects in safety-critical applications.
    \item[$\bigcdot$] \emph{Interpretability}: ML models are very complex and difficult to understand or explain, as it is reported in \cite{schindler2020real} that a recent ML model proposed for electric drive applications can have close to ten thousand parameters, not to mention those commercially deployed ML models for natural language processing or image recognition tasks that could have millions or billions of parameters. Although interpretability does not ensure safety by itself, it is important for monitoring functional safety and for understanding where the models are failing. Therefore, more in-depth investigations regarding the interpretability and explainability of ML models are necessary for their commercial deployment in drive applications.
\end{enumerate}
%
%


Upon resolving many of the practical issues mentioned above, it is anticipated that the ML-based data-driven control and monitoring schemes will be able to deliver unparalleled performance in terms of quick exploration and domain adaptation. Therefore, it has great potential to become the next-generation electric machine drive technology over the existing model-driven methods currently implemented in low-cost microcontrollers.

\balance
\bibliography{IEEEabrv.bib,ref.bib} 
%
%
\balance
%
%

\end{document}